\documentclass[fleqn,usenatbib]{mnras}

\usepackage{newtxtext,newtxmath}
\usepackage[T1]{fontenc}

\DeclareRobustCommand{\VAN}[3]{#2}
\let\VANthebibliography\thebibliography
\def\thebibliography{\DeclareRobustCommand{\VAN}[3]{##3}\VANthebibliography}

\usepackage{graphicx}	% Including figure files
\usepackage{amsmath}	% Advanced maths commands
\usepackage{xcolor}
\usepackage{ulem}
\usepackage{bm}

\newcommand{\soft}{_\mathrm{soft}}

\newcommand{\CE}{_\mathrm{CE}}
\newcommand{\init}{_\mathrm{i}}

\newcommand{\surf}{_\mathrm{surf}}

\newcommand{\jet}{_\mathrm{j}}
\newcommand{\env}{_\mathrm{e}}

\newcommand{\acc}{_\mathrm{a}}

\newcommand{\Edd}{_\mathrm{Edd}}

\newcommand{\amb}{_\mathrm{amb}}

\newcommand{\En}{\mathcal{E}} % energy density 
\newcommand{\vel}{_v}
\newcommand{\half}{_\mathrm{h}}
\newcommand{\adv}{_\mathrm{adv}}

\newcommand{\choke}{_\mathrm{c}}

\newcommand{\bind}{_\mathrm{b}}
\newcommand{\den}{_\rho}
\newcommand{\net}{_\mathrm{net}}

\newcommand{\unb}{_\mathrm{ub}}
\newcommand{\orb}{_\mathrm{orb}}
\newcommand{\scriptr}{\xi}
\newcommand{\zero}{0}

\newcommand{\gcmcmcm}{\,{\rm g\,cm^{-3}}}

\newcommand{\dyncmcm}{\,{\rm dyn\,cm^{-2}}}
\newcommand{\K}{\,\mathrm{K}}
\newcommand{\erg}{\,\mathrm{erg}}
\newcommand{\ergs}{\,\mathrm{erg\,s^{-1}}}
\newcommand{\s}{\,{\rm s}}     
\newcommand{\da}{\,{\rm d}}     
\newcommand{\yr}{\,{\rm yr}}     
\newcommand{\Rsun}{R_\odot}
\newcommand{\Msun}{M_\odot}

\newcommand{\au}{\,\mathrm{au}}
\newcommand{\km}{\,\mathrm{km}}
\newcommand{\kms}{\,\mathrm{km\,s^{-1}}}
\newcommand{\solarmassyr}{M_\odot\mathrm{\,yr^{-1}}}
\newcommand{\Msunyr}{M_\odot\mathrm{\,yr^{-1}}}
\newcommand{\luke}[1]{\textcolor{red}{#1}}

\title[Jets during common envelope evolution]{Jets from main sequence and white dwarf companions during common envelope evolution}

\author[Y. Zou et al.]{
Yangyuxin Zou,\thanks{E-mail: yzou5@ur.rochester.edu}
Luke Chamandy,\thanks{E-mail: lchamandy@pas.rochester.edu}
Jonathan Carroll-Nellenback,
Eric G. Blackman
\newauthor
~and Adam Frank 
\\
Department of Physics and Astronomy, University of Rochester, Rochester, NY 14620, USA\\
}

\date{Accepted XXX. Received YYY; in original form ZZZ}

\pubyear{2021}

\begin{document}
\label{firstpage}
\pagerange{\pageref{firstpage}--\pageref{lastpage}}
\maketitle
% Abstract of the paper, no longer than 250 words
\begin{abstract}
It has long been speculated that jet feedback from accretion onto the companion during a common envelope (CE) event could affect the orbital evolution and envelope unbinding process. 
%,
%but this conjecture has heretofore remained largely untested.
We present global 3D hydrodynamical simulations of CE evolution (CEE) that include a jet subgrid model
and compare them with an otherwise identical model without a jet.
Our binary consists of a $2\Msun$ red giant branch primary 
and a $1\Msun$ or $0.5\Msun$ main sequence (MS) or white dwarf (WD) 
secondary companion modeled as a point particle. 
We run the simulations for 10 orbits (40 days).
Our jet model adds mass at a constant rate $\dot{M}\jet$ of order the Eddington rate, 
with maximum velocity $v\jet$ of order the escape speed,
to two spherical sectors with the jet axis perpendicular to the orbital plane. %,
%and supplies kinetic energy at the rate $\sim\dot{M}\jet v\jet^2/40$.
We explore the influence of the jet on orbital evolution, envelope morphology and envelope unbinding,
and assess the dependence of the results on jet mass-loss rate, launch speed, companion mass, opening angle,
and accretion rate.
%and whether or not subgrid accretion is turned on.
In line with our theoretical estimates, 
%we find that  in all cases the 
jets are 
%becomes 
choked around the time of first periastron passage and remain choked thereafter.
Subsequent to choking, but not before, jets efficiently transfer energy to bound envelope material.
%We also find that jets 
This leads to increases in unbound mass of up to $\sim10\%$,
as compared to the simulations without jets.
%which do not include a jet.
% \luke{
We also estimate the cumulative effects of jets over a full CE phase,
finding that jets launched by MS and WD companions are unlikely to dominate envelope unbinding. %}
%\luke{We estimate that, over a full CE event, 
%kjets launched by MS or WD companions are unlikely to dominate envelope unbinding.}
\end{abstract}

% Select between one and six entries from the list of approved keywords.
% Don't make up new ones.
\begin{keywords}
binaries: close -- stars: jets -- hydrodynamics -- 
stars: winds, out flows -- white dwarfs 
\end{keywords}

%%%%%%%%%%%%%%%%%%%%%%%%%%%%%%%%%%%%%%%%%%%%%%%%%%

%%%%%%%%%%%%%%%%% BODY OF PAPER %%%%%%%%%%%%%%%%%%

\section{Introduction}
\label{sec:intro}
The common envelope (CE) phase of binary stellar evolution \citep{Paczynski76,Ivanova+13a} is believed to play a key role
in the lead-up to several astrophysical phenomena, including certain types of planetary nebulae
and supernovae and mergers of neutron stars (NSs) or stellar mass black holes (BHs).
However, CE evolution (CEE) is still not very well understood, 
partly owing to its intrinsic 3D nature, the vast range of spatial and temporal scales involved,
and the multitude of physical effects that could be important.
In CEE, the core of the primary star and the much smaller and more compact secondary companion
transfer their mutual orbital energy and angular momentum to the envelope as they inspiral, 
reducing its  binding energy.

Energy sources other than orbital may also be important for envelope unbinding.
One possibility is potential energy that is liberated as envelope material accretes onto the secondary.
Main sequence (MS) stars in post-CE binary systems that are inflated in size \citep{Jones+15} 
or chemically enriched with carbon \citep{Miszalski+13} might be signatures of accretion during CEE. 
While CEE theoretical models differ in the treatment of  accretion -- which generally occurs at subgrid scales -- they concur that envelope mass can  flow toward the secondary at rates much higher than the photon Eddington limit $\dot{M}\Edd$ 
(\citealt{Ricker+Taam08}, \citealt{Ricker+Taam12}, \citealt{Blackman+Lucchini14}, \citealt{Macleod+17}, \citealt{Chamandy+18};
see the latter reference for a derivation of $\dot{M}\Edd$).
This suggests that some accretion at or in excess of the Eddington rate is plausible.

Part of the liberated accretion energy might be directed into bipolar jets,
which would in turn help to transport this energy to other parts of the envelope \citep[e.g.][]{Soker04}.
% \amyt{
A jet could be launched from the centre of an accretion disk and be accelerated and collimated by a hydromagnetic Blandford-Payne magneto-centrifugal launch mechanism \citep{Blandford+Payne1982} or a magnetic tower \citep{Lynden-Bell2003}. 
% \lukec{Need to reword the above, mentioning the magnetic tower model and providing a reference...}\amy{added a reference for magnetic tower, took out the reference to \citet{Livio2011} which provides a short review of astrophysical jets}
The jet in turn removes mass, angular momentum, and pressure from the accretion flow and %}
could facilitate super-Eddington accretion.
Indeed, simulations have shown that accretion discs 
 can develop around the secondary, 
supporting  the idea that jets could be launched in some cases \citep{Murguia-berthier+17,Chamandy+18}.
The secondary can also accrete material just before CEE, and thus potentially exhibit a jet upon  entering CEE
(e.g.~\citealt{Chen+17}).

Various theoretical studies have explored the role of jets launched from the secondary star during CEE.
Several  have focused on NS or BH secondaries. 
In these cases, 
photons could be trapped and advected by the accretion flow, 
but accretion energy and pressure buildup can be  released  by weakly interacting neutrinos.
So-called hypercritical accretion, at rates $\gtrsim10^4\dot{M}\Edd$, may thus be common
\citep{Chevalier93,Fryer+96,Brown+00,Armitage+Livio00,Chevalier12,Lopez-camara+19,Cruz-osorio+Rezzolla20}.
These studies have  generally employed hydrodynamic simulations with ``wind-tunnel'' setups that 
are local in  simulating only a small region around the secondary.
They are generally appropriate only for modeling the earliest stages of CEE,
before the binary separation $a$ becomes comparable to the Bondi-Hoyle-Lyttleton accretion radius
\citep{Chamandy+19b,Everson+20}.

\citet{Shiber+19} presented the first 
\textit{global} CE simulations to model jet feedback
and allow orbital evolution and compared these simulations to 
others which did not include jets.
These adaptive mesh refinement (AMR) simulations involved a $0.88\Msun$ red giant branch (RGB) star of radius $83\Rsun$ 
with a $0.39\Msun$ core (represented by a point particle) along with a $0.3\Msun$ MS secondary
(also represented by a point particle and with jet properties chosen assuming it to be an MS star).

Comparing similar simulations with and without a jet,
\citet{Shiber+19} concluded that the presence of a jet increased the outward flux of material 
through a sphere of radius $1\au$ or $2\au$ 
centred on the origin of the simulation domain and extending out to the boundary.
The fraction of the material which had positive energy density flowing out of this sphere was also 
determined to be higher when a jet was included.
Furthermore, they found that the orbits of the particles were quite different 
between jet simulations and corresponding no-jet simulations,
with jet runs settling into a larger-separation orbit by the end of the simulation compared to no-jet runs.
They also found that jets can get choked as the secondary enters denser envelope gas, 
but that they can eventually break out along the polar directions.

Recently, \citet{Lopez-camara+21} performed local (in space and time) CE accretion/jet simulations 
that read in global simulation data from simulation\#5 (without jet) of \citet{Shiber+19}.
Their accretion and jet subgrid models were more sophisticated and realistic than those of \citet{Shiber+19}
and they employed $16\times$ higher maximum resolution.
\citet{Lopez-camara+21} found that jets are choked before the first periastron passage, earlier than in \citet{Shiber+19}.
Moreover, whereas \citet{Shiber+19} found that a jet can break out again after $\sim3\da$, 
\citet{Lopez-camara+21} found that the jet had still not broken out $\sim10\da$ after choking, when their simulations ended.

Aside from the 3D simulation studies mentioned above,
there have been other theoretical studies finding that jets can be important for envelope unbinding.
It has been suggested that CEE can be avoided if the system instead enters a quasi-steady 
``grazing envelope evolution'' (GEE) phase,
where the jet of the secondary continuously removes material from the outskirts of the envelope \citep{Soker15}. 
However, GEE has so far not been shown to take place in a 3D hydrodynamical CE simulation 
that includes all relevant gravitational interactions.
Another possibility is that a jet launched by the secondary 
could help to unbind what remains of the envelope during the late stages of CEE \citep{Soker17b}.

The primary goal of the present work is to investigate to what degree a jet launched from the secondary
can affect envelope unbinding during CEE, and we also study how jets affect
orbital evolution and envelope morphology. 
The organization of the paper is as follows. In Section~\ref{sec:methods}, we describe the numerical methods and model. 
Then in Section~\ref{sec:theory} we present some order-of-magnitude analytical estimates 
that can assist in the interpretation of numerical results 
and also allow for extrapolation to regions of the parameter space inaccessible in the simulations.
In Section~\ref{sec:results}, we present the results of our simulations.
We discuss the implications of our results for envelope unbinding in Section~\ref{sec:disc_unb},
and compare our findings to the literature in Section~\ref{sec:literature}.
In Section~\ref{sec:conclusions} we summarize our key results and conclude.

\section{Methods}\label{sec:methods}
Hydrodynamic simulations in this work were carried out using the AMR code \textsc{astrobear}, 
\citep{Cunningham+09, Carroll-nellenback+13}.
The simulation setups are very similar to those in our previous works \citep{Chamandy+18,Chamandy+19b} to which we now add a jet.
We use an ideal gas equation of state with adiabatic index $\gamma=5/3$.
The primary is an RGB star of mass $1.96\Msun$ and radius $48.1\Rsun$,
and its core is modeled using a $0.367\Msun$ point particle
according to the method of \citet{Ohlmann+17}, along with an added iteration over particle mass \citep{Chamandy+18}.
For most of our runs the secondary is a point particle with initial mass $0.978\Msun$, 
but for two runs we use $0.489\Msun$ instead.
The spline softening radius for the primary core and secondary particles is $2.41\Rsun$ 
and the simulations employ $4$ levels of AMR with 
base resolution $\delta_0=2.25\Rsun$ and highest resolution $\delta_4=0.14\Rsun$.
The density and pressure of the ambient medium surrounding the star are $6.7\times10^{-9}\gcmcmcm$ and $1.0\times10^5\dyncmcm$,
and the cubic domain has side length $1150\Rsun$.
The simulation is conducted in a reference frame in which the centre of mass remains approximately fixed. 
The two stars are initialized in a circular orbit at $t=0$, with $a\init = 49\Rsun$,
and the envelope has no initial rotation.  

\subsection{Jet model}
\label{sec:jet_model}
We use  a slightly modified version of the jet subgrid model of \citet{Federrath+14}, 
with a scheme that conserves mass, momentum and angular momentum while minimizing discretization errors, 
as detailed in Appendix~\ref{sec:jet_details}.
%Parameters of the subgrid jet model are kept constant in time for simplicity.
For the duration of the simulations, gas is added to the grid at the rate $\dot{M}\jet$,
divided between two oppositely oriented spherical sectors centred around the companion particle. 
The jet axis is constrained to be perpendicular to the orbital plane.
Each sector extends from the vicinity of the companion out along the jet axis 
to the distance $r\jet = \delta_0 = 16\delta_4 = 2.25\Rsun$,
and from the axis to the polar angle $\theta\half$ (the jet half-opening angle).
The jet density and radial velocity fall off at higher polar angle $\theta$, 
and the jet velocity is strongly peaked near the jet axis within $\theta<\theta\half/6$.
Further, the jet mass is concentrated around $r\sim r\jet/2$
(see \citealt{Federrath+14} for details of the jet geometry). 
In addition, the jet material is initialized with an extra velocity component 
equal to the instantaneous orbital velocity of the secondary.
Owing to the \citet{Federrath+14} density and velocity profiles chosen for the jet,
the rate of kinetic energy supplied to the jet (excluding the smaller variable contribution from the orbital motion) 
is equal to about $\dot{M}\jet v\jet^2/40$,
with $v\jet$ the peak outward velocity of the jet, which occurs at $\theta=0$.
The jet temperature is chosen to be $T\jet=10^4\K$, 
but the results are insensitive to temperature because the jet is highly supersonic.
The jet is initialized to be non-rotating.

%\luke{
The jet model we use \citep{Federrath+14} is structured to include a high momentum spine with small opening angle 
and a lower momentum ``wide angle wind''.  
This structured radial momentum distribution in the jet is in contrast to simple constant density, 
constant velocity ``top hat'' jets.  
While simpler, top hat jets are not as realistic since it is expected that collimated flows are driven 
by magnetocentrifugal processes from the star/disc system.  
Both analytic models \citep{Shu+00} and numerical simulations \citep{Banerjee+Pudritz06, Machida+08} 
of such systems show that the outflows they produce always show 
%\luke{
momentum
%}
%density 
falling off sharply as one moves away from the axis of the flow. 
%}
%\lukec{I replaced density with momentum. Is this okay?}

The initial jet speed at the stellar surface
may be expressed as $Q$ times the Keplerian value, 
i.e.~$v\surf=432Q\kms\,(M_2/M_\odot)^{1/2}(R_2/\Rsun)^{-1/2}$,
with $Q>1$ \citep[e.g.][]{Blackman+Lucchini14}.  
The value of the peak speed $v\jet$, at the jet initialization radius of $r\sim r\jet/2\approx1\Rsun$,
is then chosen by setting $Q$ to some value between $2$ and $4$.
%To fix the jet initialization \luke{peak} speed $v\jet$, 
For models for which the secondary point particle represents a WD ($R_2\sim0.01\Rsun$), 
we also take into account the influence of gravity from the secondary when choosing $v\jet$.
%\lukec{I rearranged the text in this paragraph to improve clarity and remove unnecessary details.}

% \erict{
Our jet prescription is not intended to capture the detailed mechanism of any specific jet model  such as  magneto-centrifugal launch (MCL) models 
\citep{Blandford+Payne1982,Pelletier+1992} or 
magnetic tower (MT) \citep{Lynden-Bell2003, Huarte-espinosa+13}. We merely specify the kinematics and opening angle.  However, because we specify the kinematics of the jet to be dominated by a velocity flow, our specification is most consistent asymptotically with  MCL type models rather than  MT models because the latter can remain magnetically dominated out to larger distances from the engine. MCL models are   asymptotic hydrodynamically dominated, and so our prescription is essentially  taking the asymptotic properties of the MCL as the input. %} 

In our model the secondary launches a jet with constant $\dot{M}\jet$ and %\luke{
peak velocity %}
$v\jet$ from time $t=0$.
This effectively assumes that the (subgrid) processes governing accretion and jet launching
are in a steady state that is unaffected by the changing environment of the secondary over the timescale of the simulation.  Aside from simplifying the numerical implementation, 
this choice minimizes the number of variables, 
thus facilitating interpretation of the simulation results.

For most of our runs, the secondary is not permitted to accrete from the surrounding envelope.
For some runs we allow it to accrete at roughly the  Bondi-Hoyle-Lyttleton rate,
as computed by our accretion subgrid scheme, which is modeled after \citet{Krumholz+04} (see also \citealt{Chamandy+18}).
Unlike \citet{Krumholz+04} our scheme allows the point particle secondary to remove angular momentum from the gas it accretes.
The specific angular momentum of gas inside the accretion zones that is \textit{not} accreted
is conserved during the accretion step, as in the scheme of \citet{Federrath+10}.
For our WD run, we cap the accretion rate at $\approx10^4\dot{M}\jet$.
As radiative feedback is not included, the accretion rates that obtain can be considered upper limits.

Separate tracers are applied to the jet, envelope and ambient,
which allows us to distinguish these components in post-processing.
Note that co-spatial jet gas and envelope/ambient gas have different densities, but the same velocity and temperature.
Simulations are carried out for $40\da$, or about $10$ orbits in most of our models;
this duration was chosen to optimize computational resources.

\begin{table}
  \centering
  \setlength\tabcolsep{3pt} % default value: 6pt
  \begin{tabular}{@{}l c c c c c c c @{}}
    \hline
Model  &Type    &$M_2(t=0)$    &$\dot{M}\jet$      &$\dot{M}\acc$ &$\dot{M}_2$               &$v\jet$    &$\theta\half$  \\
       &        &[$0.978\Msun$]&[$10^{-3}\Msunyr$] &              &                          &[$\!\kms$] &[$^\circ$]     \\
\hline                      
J1     &MS      &$1$           &$2$                &$0$           &$-\dot{M}\jet$            &$864$      &$15$           \\
NJ1    &---     &$1$           &$0$                &$0$           &$0$                       &---        &---            \\
J2     &MS      &$0.5$         &$20$               &C18           &$\dot{M}\acc-\dot{M}\jet$ &$864$      &$15$           \\
NJ2    &---     &$0.5$         &$0$                &$0$           &$0$                       &---        &---            \\
J3     &MS      &$1$           &$2$                &$0$           &$0$                       &$864$      &$15$           \\
J4     &MS      &$1$           &$2$                &C18           &$\dot{M}\acc-\dot{M}\jet$ &$864$      &$15$           \\
J5     &MS      &$1$           &$20$               &C18           &$\dot{M}\acc-\dot{M}\jet$ &$864$      &$15$           \\
J6     &MS      &$1$           &$2$                &$0$           &$-\dot{M}\jet$            &$1728$     &$15$           \\
J7     &MS      &$1$           &$2$                &$0$           &$-\dot{M}\jet$            &$864$      &$30$           \\
J8     &WD      &$1$           &$0.02$             &C18           &$\dot{M}\acc-\dot{M}\jet$ &$8640$     &$15$           \\
\hline
\end{tabular}
\caption{Models labeled with `J' refer to runs with a jet, and those labeled with `NJ' refer to runs without a jet. 
         Jet parameters were selected based on whether the secondary is modeled as a main sequence (MS) star, 
         or a white dwarf (WD).
         %\luke{
         The quantity $v\jet$ is the peak jet velocity. %}
         For all runs, $r\soft=2.41\Rsun$, $r\jet=\delta_0=2.25\Rsun$ and $T\jet=10^4\K$.
         The Eddington accretion rate $\dot{M}\Edd\sim2\times10^{-3}(R_2/R_\odot)~\solarmassyr$ (e.g.~\citealt{Chamandy+18},
         which is denoted as `C18' in the table).
         }
\label{tab:runs}
\end{table}

\subsection{Runs}
\label{sec:runs}
Table~\ref{tab:runs} describes the runs performed.
J1 is the fiducial jet run and NJ1 is the fiducial no-jet ($\dot{M}\jet=0$) run.
In nature, we might expect $\dot{M}\jet/\dot{M}\acc\sim0.1$, where $\dot{M}\acc$ is the accretion rate,
and $\dot{M}\acc/\dot{M}\Edd\lesssim10$ with the upper bound $>1$ 
because the Eddington limit can be modestly exceeded in non-spherical flows.
For most of our runs, $\dot{M}\jet$ is taken to equal the estimated Eddington rate
for a $1\Rsun$ MS star, $\dot{M}\jet=\dot{M}\Edd\approx2\times10^{-3}\Msunyr$ \citep[e.g.][]{Chamandy+18}.
In some cases we adopt the more extreme value $\dot{M}\jet\approx10\dot{M}\Edd$. 
%However, because of the jet density and velocity profiles employed (see Section~\ref{sec:methods}),
%the kinetic energy of the jet is only about $1/20$ of what a jet with \luke{a} uniform velocity \luke{of} $v\jet$ would have.
%\eric{the above sentence is confusing to the reader so I commented it out.  Can we  simply change the label of the velocity column to simply be "peak" velocity and in the table caption?}
%\lukec{Agree with your edit and slightly tweaked sentence as well just now. 
%The table contains the parameter symbols so it would be a bit out of place to replace with words.
%Also, the table would no longer fit in a column (without further changes).
%I added a sentence to the table caption. Is this good enough?}
%Moreover, the \textit{effective} half-opening angle of the jet is $\theta\half/6$, 
%as the kinetic energy is concentrated along the central axis.
%\lukec{Modified this paragraph so as not to belabor the point about jet profile...}

For most of our models, the secondary loses mass at the rate $\dot{M}_2=-\dot{M}\jet$ to conserve mass and momentum.
The secondary along with the gas inside the softening sphere 
loosely represent a star+accretion disc \textit{system}
with small-scale (subgrid) processes enforcing a fixed value of $\dot{M}\jet$.
To test the sensitivity to this aspect of the modeling, we perform Run J3, 
which is the same as J1 except that $M_2$ is fixed.
Furthermore, 
%\luke{
as mentioned above,
%} 
for some of our models the secondary 
%\luke{
point particle
%} 
is permitted to accrete from its surroundings at
very high rates (up to $\sim10^3\dot{M}\Edd$ for our MS star runs and $\sim10^4\dot{M}\Edd$ for our WD run).
These accretion rates are upper limits,
and allow us to explore qualitatively the effects of accretion by comparing to those runs without accretion.
For example, Run J4 is like J1 except that it allows accretion.

To test the dependence of the results on the parameters, 
we have Runs J6 and J7, which are like J1 except with $v\jet$ or $\theta\half$ doubled.
Run J5 is like J4 but with a 10 times higher $\dot{M}\jet$.
Run J2 (corresponding no-jet run NJ2) is like J5 except that the secondary mass is halved. 
Finally, in Run J8 we adopt jet parameters that are appropriate for a WD companion.
J8 is like J4 but with a 10 times larger jet speed and 100 times smaller jet mass-loss rate,
resulting in a jet with kinetic energy input rate comparable to that of J4 or J1.

\section{Theoretical estimates}
\label{sec:theory}

\subsection{Jet launching}

\label{sec:launch}
From $t=0$, material is added to the jet spherical sectors at the rate $\dot{M}\jet$, 
with vertical speed 
%\luke{
of order
%} 
$v\jet$.
This added material does not propagate out of the jet sectors immediately owing to the inertia of the ambient gas.
The material in the jet sectors is accelerated to the initialization velocity after a time $t\vel$,
when sufficient momentum has been imparted.
We crudely estimate this time 
neglecting the variation in the density and  vertical component of the velocity across the jet. 
At $t=0$, when the pressure and gravity of the ambient gas are negligible,
we have
\begin{equation}
  \label{accel}
  \frac{\dot{M}\jet v\jet}{2}\sim \rho V\frac{v\jet}{t\vel},
\end{equation}
where the factor of $2$ accounts for the two jet spherical sectors,
$V= 2\uppi r\jet^3\left(1-\cos\theta\half\right)/3$ is the volume of each sector, and $\rho=\rho\amb$.
Solving for $t\vel$ we obtain
\begin{equation}
\label{tv}
  t\vel\sim \frac{4\uppi \rho r\jet^3(1-\cos\theta\half)}{3\dot{M}\jet },
\end{equation}
which evaluates to $t\vel\sim30\s$ for our fiducial model.
At this time the material in the spherical sectors moves at a speed 
%\luke{
of order
%} 
$v\jet$ 
and still has a density comparable to $\rho\amb$.

Note that $\rho\amb$ is lower than the anticipated final density 
$\rho\jet\sim\dot{M}\jet /(2Av\jet)\sim5\times10^{-7}\gcmcmcm$,
where $A\sim\uppi(r\jet/2)^2\tan^2\theta\half$ is the jet cross-sectional area
and the numerical estimate is for our fiducial model.
The jet is truly launched when material is advected out of the spherical sectors, 
which happens after a time $t\adv\sim r\jet/v\jet\sim2\times10^3\s$ for our fiducial model.
This is also approximately equal to the time it takes for the density in the spherical sectors to reach $\rho\jet$,
given by $t\den\sim 2V\rho\jet/\dot{M}\jet \sim V/Av\jet\sim r\jet/v\jet$.

\subsection{Jet choking}
\label{sec:choke}
As the secondary plunges deeper into the envelope,
the ram pressure force of the jet on  the surrounding gas 
remains roughly constant but the gravitational force due to the secondary increases because of the higher density.
As the jet velocity and mass flux density in our model 
are strongly peaked within $\theta<\theta\half/6$ \citep{Federrath+14},
we focus on the choking of this central part of the jet.
The jet is choked when the net force reduces to zero,
\begin{equation}
  \label{force}
  F\net\sim\frac{(\dot{M}\jet/5) v\jet}{2}-\frac{GM_2\rho\choke (V/36)}{(r\jet/2)^2}=0,
\end{equation}
where $\rho\choke$ is the critical density just outside the jet launch region.
Here $\dot{M}\jet$ is divided by $5$ since we are considering only the fraction of the mass flux 
associated with the central part of the jet,
$\int_0^{\theta\half/6}\rho(r,\theta)v(\theta)\sin\theta d\theta 
\Big/\int_0^{\theta\half}\rho(r,\theta)v(\theta)\sin\theta d\theta\approx0.2$, 
and $V$ is divided by $36$ since the central part occupies the portion 
$[1-\cos(\theta\half/6)]/(1-\cos\theta\half)\approx1/36$ of the total volume.
Solving for $\rho\choke$, we obtain
\begin{equation}
  \label{rho_choke}
  \rho\choke\sim\frac{0.4\dot{M}\jet v\jet}{GM_2 r\jet(1-\cos\theta\half)},
\end{equation}
which evaluates to $\rho\choke\approx7\times10^{-6}\gcmcmcm$ for our fiducial parameter values.
In the initial envelope, this density occurs at a radius of about $37\Rsun$,
so we estimate that the jet could choke at this separation, $a\choke\sim37\Rsun$.
In an identical simulation for which a jet is \textit{not} launched, 
this separation is reached after $t\choke\sim9\da$ \citep[][or Model~NJ1]{Chamandy+18}.
We do not expect the small mass loss rate of the jet to have a large effect on the orbit at this time
(this is confirmed by our numerical results, presented below).
Therefore, we might expect the jet to choke at about this time.

Additionally,
we can use equation~\eqref{tv} with $\rho=\rho\choke$
to find the time it would take to accelerate envelope material in the jet launch region up to $\luke{\sim}v\jet$,
now assuming that the jet ram pressure force still dominates over gravity.
Since $\rho\choke/\rho\amb\sim10^3$ we obtain $t_\mathrm{v,q}\sim3\times10^4\s$ or about $0.3\da$.
This can be compared with the time it takes for new envelope material to enter the jet launch region
\begin{equation}
  \label{torb}
  t\orb\sim\frac{2r\jet\tan\theta\half}{v\orb}, 
\end{equation}
where $v\orb$ is the orbital velocity at $t=9\da$. 
Computing the tangential component of the relative velocity 
of the secondary with respect to the primary core particle from the simulation,
we obtain $v\orb\approx100\kms$.
This gives $t\orb\sim0.1\da$, which is $\lesssim t_\mathrm{v,q}$,
so the jet has become unable to clear away envelope material before new material takes its place. 
%In reality, the gravity of the companion increases the surrounding envelope density and pressure
%compared to that in the initial envelope \citep{Chamandy+19b}.
%Therefore, our estimate of the choking time here can be taken as an upper limit.
% \luke{
In reality the secondary alters the local conditions in the envelope,
so these estimates are only rough.
% }
% \lukec{Removed some unnecessary speculation.}

We conclude that jet choking 
%likely commences 
% \luke{
is expected to commence
% }
%before 
% \luke{
by
% } 
the first periastron passage. 
This conclusion applies to our other runs as well.

\subsection{Role of the jet in unbinding the envelope}
\label{sec:unbinding1}
\subsubsection{Overall contribution during the CE phase}
\label{sec:tjet}

Even if the jet does not contribute significantly to envelope unbinding over the $10$-orbit timescale of our simulations,
it may  be significant over the longer full CE phase,
whose termination is presumably marked by stabilization of the orbit and complete ejection of the envelope.

The timescale for a jet from the secondary to supply energy comparable to the binding energy of the envelope
is $E\bind/\dot{E}\jet$.
Here $E\bind=1.9\times10^{47}\erg$ is the magnitude of the binding energy of the envelope, 
including contributions from thermal energy and potential energy terms involving the envelope self-gravity
and the gravitational interaction between the primary core point particle and envelope.

The jet is expected to be choked early on (Section~\ref{sec:choke}).
% \luke{
Consider a more extreme case than those simulated where the jet
% }
injects energy into the envelope at a constant rate of 
$\dot{E}\jet\sim\tfrac{1}{2}\dot{M}\jet v\jet^2$ thereafter.
For the $\dot{M}\jet$ and $v\jet$ used in Runs J1 or J8, 
this gives $\dot{E}\jet\sim5\times10^{38}\ergs$,
which implies $E\bind/\dot{E}\jet\sim13\yr$,
whereas for 
%our most optimistic models with 
$10$ times higher $\dot{M}\jet$,
we obtain $E\bind/\dot{E}\jet\sim1.3\yr$.

This is the time it would take for 
%the 
% \luke{
a very powerful, continuously active
% } 
jet to unbind the envelope 
-- acting alone without any other sources of energy --
if jet energy could be transferred with maximal efficiency
$\alpha\jet\equiv \dot{E}\bind/\dot{E}\jet=1$, where $\dot{E}\bind$ 
is the rate at which energy is transferred to (bound) envelope material.
In nature and simulations, $\alpha\jet<1$ because a portion of the jet energy 
is transferred to already \textit{unbound} (former) envelope gas and a portion could directly leave the envelope.%
\footnote{In nature, radiative cooling of envelope gas would also reduce $\alpha\jet$ 
(but our simulations assume an adiabatic ideal gas, so that inefficiency is not accounted for).}
The time for the jet to unbind the envelope assuming a constant $\alpha\jet$ is
\begin{equation}
  t\jet=\frac{E\bind}{\alpha\jet\dot{E}\jet}.
\end{equation}
If $\alpha\jet=0.25$, for example, then we obtain $t\jet\approx50\yr$ if $\dot{M}\jet\sim\dot{M}\Edd$.
(In our simulations the jet adds kinetic energy at a rate about $20$ times smaller due to the jet
profile chosen, so $t\jet$ would be predicted to be $20$ times longer if the full CE phase could be simulated.)

Crudely extrapolating the mass unbinding seen in simulations of an almost identical system to the one considered here
(but without accretion or jets) leads to estimated unbinding times in the range $t\CE\sim1$-$10\yr$,
where the lower value includes the recombination and thermal energies in determining which gas is unbound,
and the higher value does not \citep{Prust+Chang19}.
Extrapolations of CE simulations involving asymptotic giant branch (AGB) primaries
suggest it might take of order $t\CE\sim10\yr$ to unbind the envelope \citep{Chamandy+20,Sand+20}.
%If $t\jet\gg t\CE$ then jets are unlikely to play a significant role in unbinding the envelope.
%Simulation results \luke{presented below} suggest that $t\jet>t\CE$ by at least a factor of a few,
%but there may be regions of parameter space for which $t\jet\approx t\CE$.
% \lukec{Commented out mention of simulation results here...
% less confusing if we keep the discussion limited to the analytics for now.
% Also commented out unnecessary discussion just above that and in the next paragraph.}
%\eric{good}

%Jets could contribute relatively more to unbinding toward the end of the CE phase
%if the rate of orbital energy injection declines significantly relative to $\dot{E}\jet$.
%Moreover, the CE phase could last for $100$ to $1000$ years in some cases, 
%but it seems unlikely that $\dot{M}\jet$ could continuously be sustained at close to the Eddington rate.
Taken as a whole, 
% \luke{
the above estimate suggests that jets from MS and WD secondaries 
could contribute significantly to unbinding for sustained jet mass-loss rates
equal to or exceeding the Eddington value, 
but could not dominate envelope unbinding.
% }
%it seems to us 
%\luke{these analytic estimates suggest it is}
%unlikely that jets from MS and WD secondaries
%%would 
%\luke{could} dominate envelope unbinding,
%%though 
%\luke{but such jets} 
%%they 
%%likely
%\luke{might}
%contribute significantly \luke{to unbinding if their}
%%jets with 
%\luke{sustained} mass-loss rates \luke{were}
%%approaching the 
%\luke{of order the} Eddington value.

%\subsubsection{Contribution to unbinding on the timescale of our simulations}
\subsubsection{Expected extra unbinding in simulations with jets}
\label{sec:energy}
We can try to predict the relative importance of jets 
% \luke{
to envelope unbinding
by comparing the energy input from jets with that from orbital decay during the simulation.
% }.
In simulations of the binary system considered here without accretion or a jet, 
only $\sim10\%$ of the envelope is unbound in the first $40\da$
(\citealt{Ohlmann+16a,Chamandy+19a,Prust+Chang19} and see Section~\ref{sec:unbinding}).
%In our simulations, 
%envelope unbinding is powered by released orbital energy and, if a jet is included, jet kinetic energy.
%Thus we can  

We can 
% \luke{
roughly
% } 
estimate the orbital energy at first periastron passage for a companion mass of $0.978\Msun$
($t\approx12.8\da$)  as $-Gm_2M_\mathrm{1,int}/2a\approx-1.06\times10^{47}\erg$, 
where $a\approx14\Rsun$ is the separation and $M_\mathrm{1,int}\approx0.8\Msun$ 
is the mass of the unperturbed primary inside $r=a$.
Subtracting this from the initial orbital energy
$-Gm_2M_1/2a\init\approx-7.4\times10^{46}\erg$
gives $\sim3\times10^{46}\erg$ of orbital energy released up to the first periastron passage.
By comparison, the jet would inject 
only $3\times10^{44}\erg$ of kinetic energy in our most extreme model (Run~J5) during this time.%
\footnote{To obtain this estimate we are ignoring the kinetic energy due to the orbital motion of the jet, 
which is somewhat smaller than that due to the outward motion of jet material.}
At the end of the simulation without accretion or jet at $t=40\da$, $a\approx7\Rsun$ and $M_\mathrm{1,int}\approx0.5\Msun$.
From $t=12.8\da$ to $40\da$, the orbital energy liberated is $\sim6\times10^{46}\erg$, 
whereas the jet in J5 injects about $6\times10^{44}\erg$ of kinetic energy.
Therefore, the rate of kinetic energy supplied by the jet amounts to only $\sim1\%$ of the change in orbital energy
for Run~J5, and $\sim0.1\%$ for Run~J1 or J8.

%But 
However, the efficiency with which released energy is used to unbind envelope mass may change when a jet is present,
% \luke{
and in Section~\ref{sec:unbinding} 
we will see that the impact of jets on envelope unbinding in the simulations 
is, in fact, more substantial than the above estimate suggests.
% }
%In Section~\ref{sec:unbinding} we show that the long term influence of a jet on envelope unbinding can 
%\luke{sometimes} be \luke{more} substantial \luke{than the above estimate suggests}.
%%\eric{tweaked above to avoid taking the reader on a wayward path} 
%\lukec{I tried to clean this up and also ``long term'' may not be appropriate here as we we a still talking about the $10$ day simulations. The extra unbinding is seen in the latter half of the simulations, but no need to dwell on that here I think.}
%\eric{ok}

\section{Simulation Results}
\label{sec:results}

\subsection{Jet evolution}
\label{sec:evolution}
We first verified from the simulations that the jet is activated as discussed in Section~\ref{sec:launch}.
The timescales for the 
% \luke{
central part of the
% } 
jet to accelerate to $v\jet$ and for the jet material 
to be advected out of the spherical sectors are roughly as predicted.
\subsubsection{Jet choking}
\label{sec:morphology}

\begin{figure}
    \centering
    \includegraphics{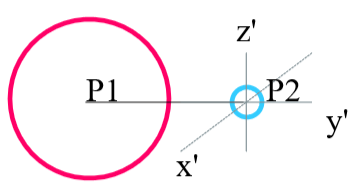}
    \caption{Diagram of the coordinate system used in this paper. The origin of this primed system is located at the point particle representing the secondary ($P2$) and the axes rotate with the binary system. The $z'$ axis aligns with the vertical 
    %direction 
    axis of the simulation lab frame $z$. The $y'$ axis is in the orbital plane of the binary and on a line joining the primary core ($P1$) and the secondary particle. The $x'$ axis is also in the orbital plane and orthogonal to the $y'$ axis. Later, we show snapshots taking slices that are orthogonal to the orbital plane, and either as viewed from the position of the primary (the $x'-z'$ plane) or sliced through both particles (the $y'-z'$ plane). }
    \label{fig:coordinate}
\end{figure}

\begin{figure*}
    \centering
    \includegraphics[width = \textwidth]{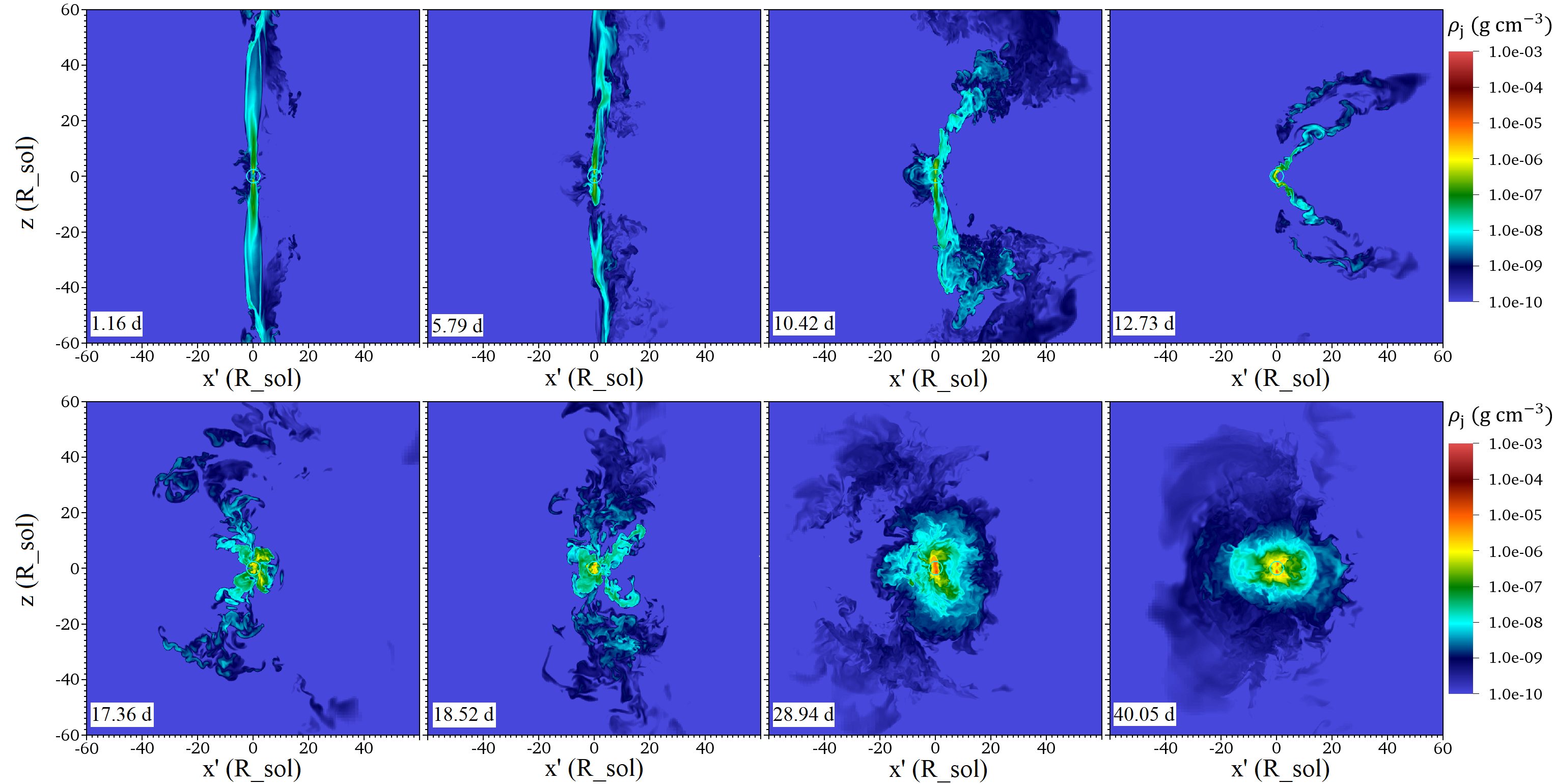}\\
    \includegraphics[width = \textwidth]{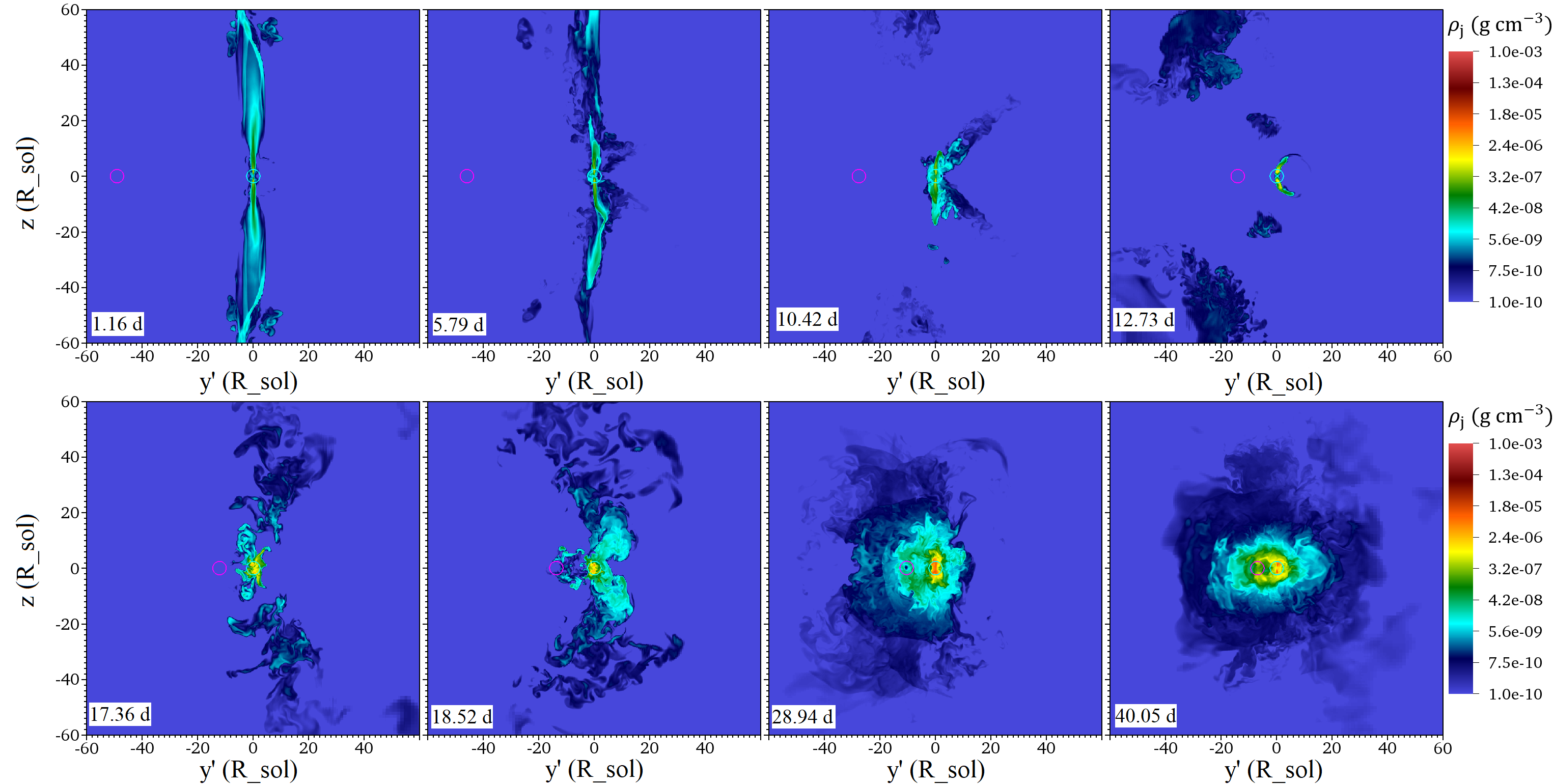}
    \caption{Snapshots showing the density of jet gas (i.e.~jet tracer) in $\gcmcmcm$
    at various times throughout the simulation model J1. 
    The secondary is located at the centre with its softening radius shown by a cyan circle.
    The jet is initially vertical and extended but becomes deformed and eventually choked inside dense envelope gas.
    In the top two rows, slices are orthogonal to the orbital plane and to the line connecting the primary core and the secondary particles (the $x'-z'$ plane),
    shown as would be viewed from the position of the primary core particle 
    (the azimuthal component of the secondary's orbital motion is toward the left). 
    The bottom two rows show the same times, but now sliced through both particles, 
    perpendicular to the orbital plane (the $y'-z'$ plane), 
    with the primary core particle situated left-of-centre.
    Its softening sphere is shown by a magenta circle.
    \label{fig:rhojet}}
\end{figure*}

\begin{figure*}
    \centering
    \includegraphics[width=\textwidth]{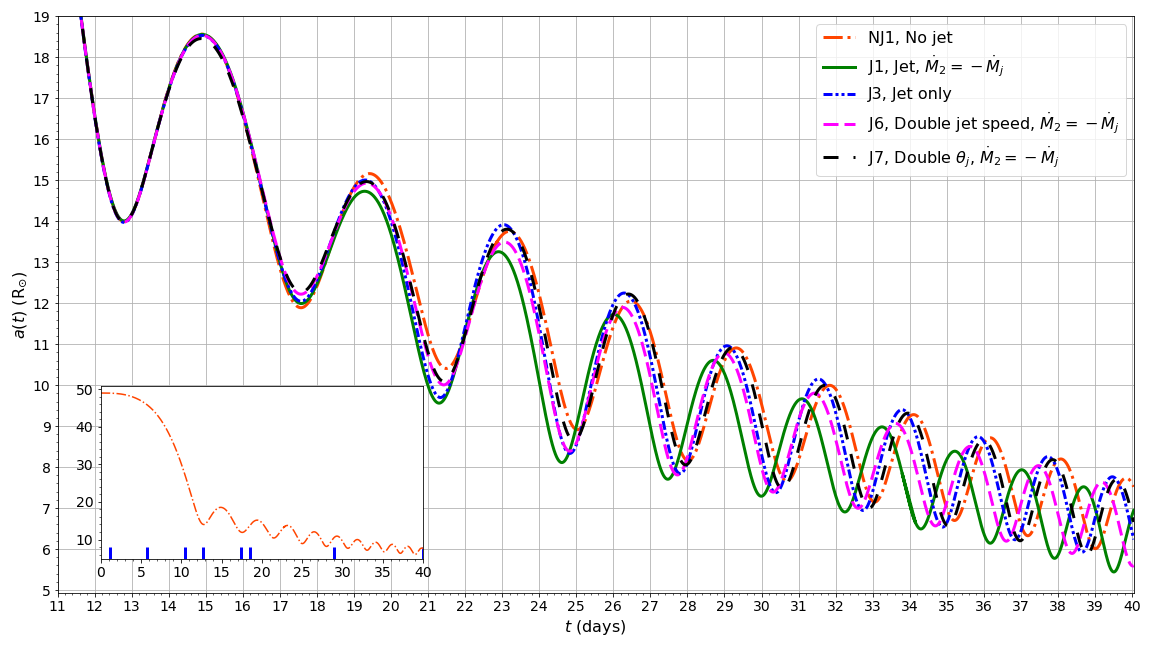} % ,clip=true,trim= 100 60 100 70
    \includegraphics[width=\textwidth]{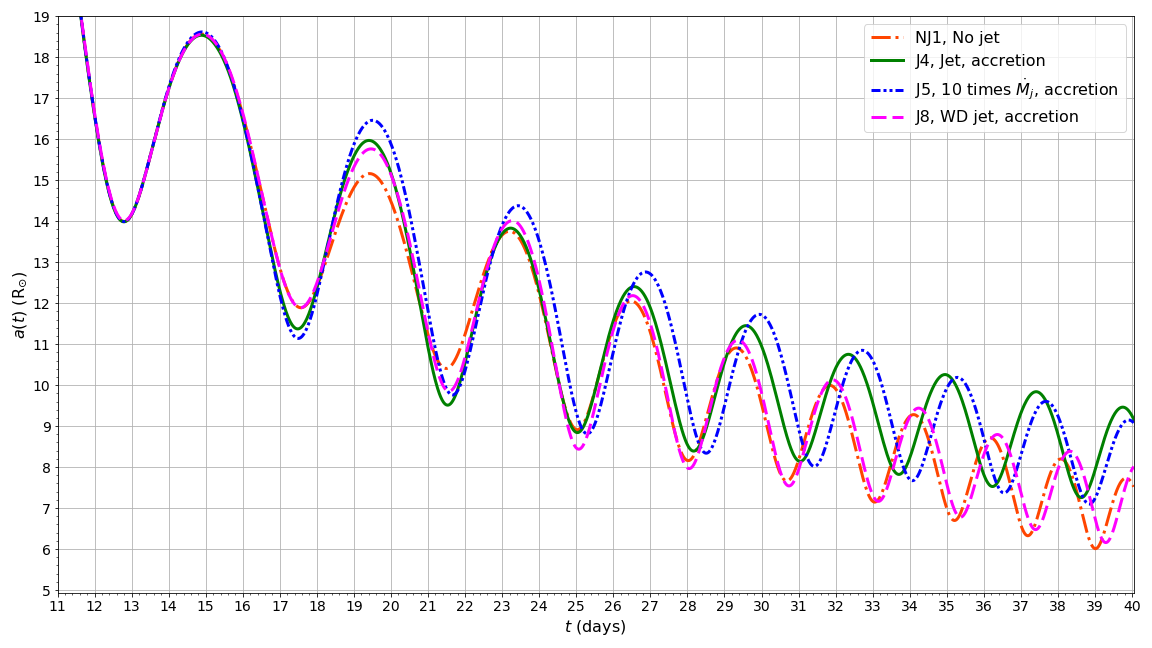} % ,clip=true,trim= 100 20 100 70
    \caption{Evolution of the inter-particle separation in the orbital plane. 
    \textit{Top:} Runs with initial companion mass $M_2(0)=0.978$ and without subgrid accretion.
    Inset shows the full time evolution of the no-jet model NJ1,
    with snapshot times of Fig.~\ref{fig:rhojet} indicated by short blue lines.
    While the orbits of all runs are almost identical up to the first periastron passage,
    all jet runs without subgrid accretion eventually develop a shorter orbital period compared to NJ1.
    \textit{Bottom:} Runs with $M_2(0)=0.978$ that include subgrid accretion, except for NJ1, 
    which is repeated from the top panel for ease of comparison.
    All jet runs with subgrid accretion eventually develop a longer orbital period than NJ1.
    }
    \label{fig:separation}
\end{figure*}

\begin{figure*}
    \centering
    \includegraphics[width=\textwidth]{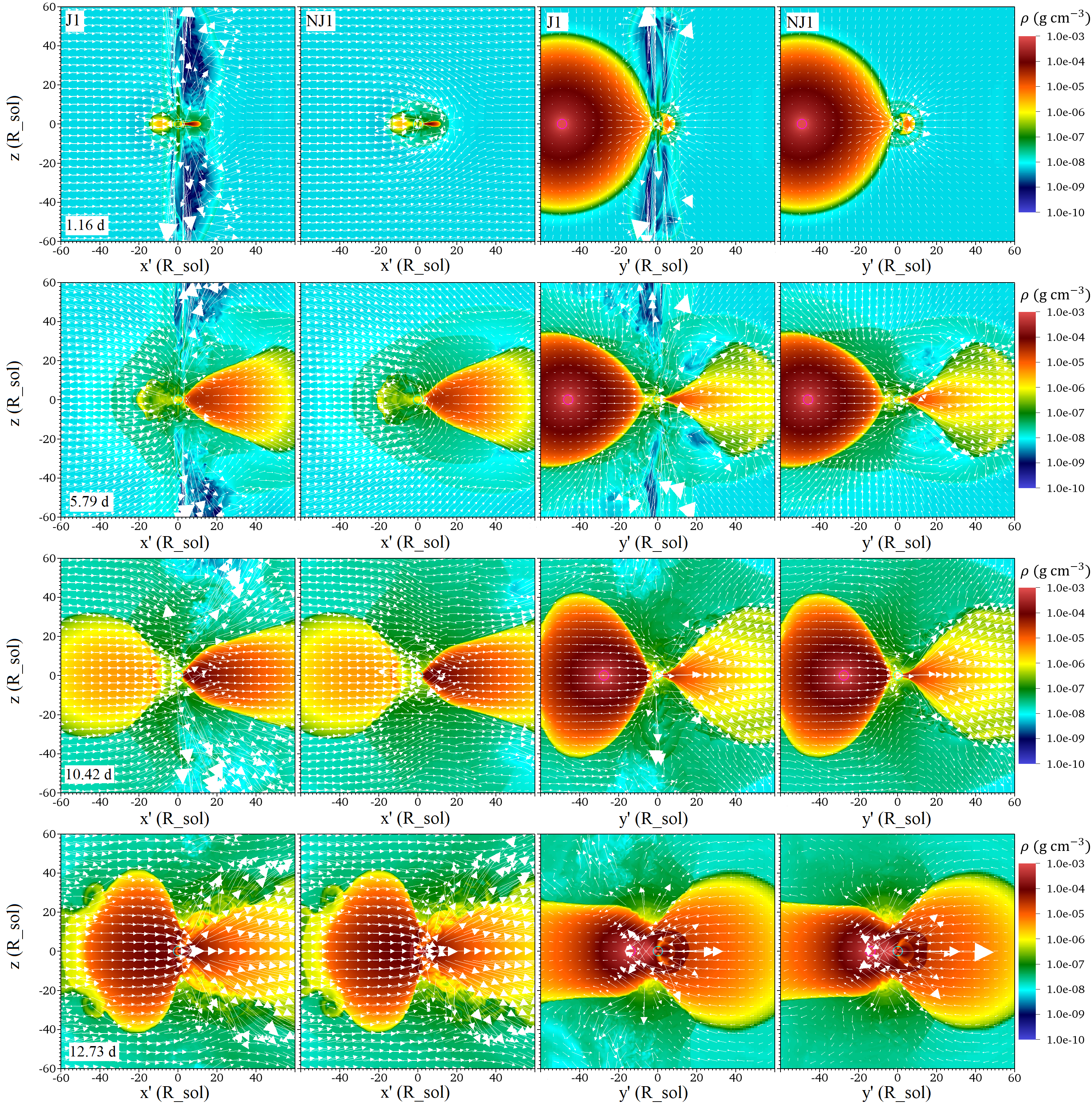}
    \caption{Snapshots of total gas density (including jet, envelope and ambient gas) comparing models J1 and NJ1.
    Times are the same as in the first four snapshots of Fig.~\ref{fig:rhojet}.
    Velocity vectors are relative to the motion of the secondary. 
    The left two columns are sliced through the secondary perpendicular to both the orbital plane
    and the line joining the particles, (the $x'-z'$ plane). %  as in Fig.~\ref{fig:rhojet}.
    The left-most column shows our fiducial jet run J1, 
    and the second column shows NJ1, which is the same except without a jet.
    By the last snapshot, at approximately the first periastron passage (bottom row),
    the differences between the jet and no-jet slices are very small.
    The right two columns show the slice through both particles and perpendicular to the orbital plane, 
    with NJ1 in the right-most column and J1 to its left (the $y'-z'$ plane).
    The softening sphere of the RGB core is shown as a magenta circle. 
    Note the bow shocks below the secondary at $t=10.42\da$ (third row, third panel from the left),
    which are produced by the jet as it chokes.
    }
    \label{fig:compare-J1-NJ1}
\end{figure*}

\begin{figure*}
    \centering
    \includegraphics[width=0.88\textwidth]{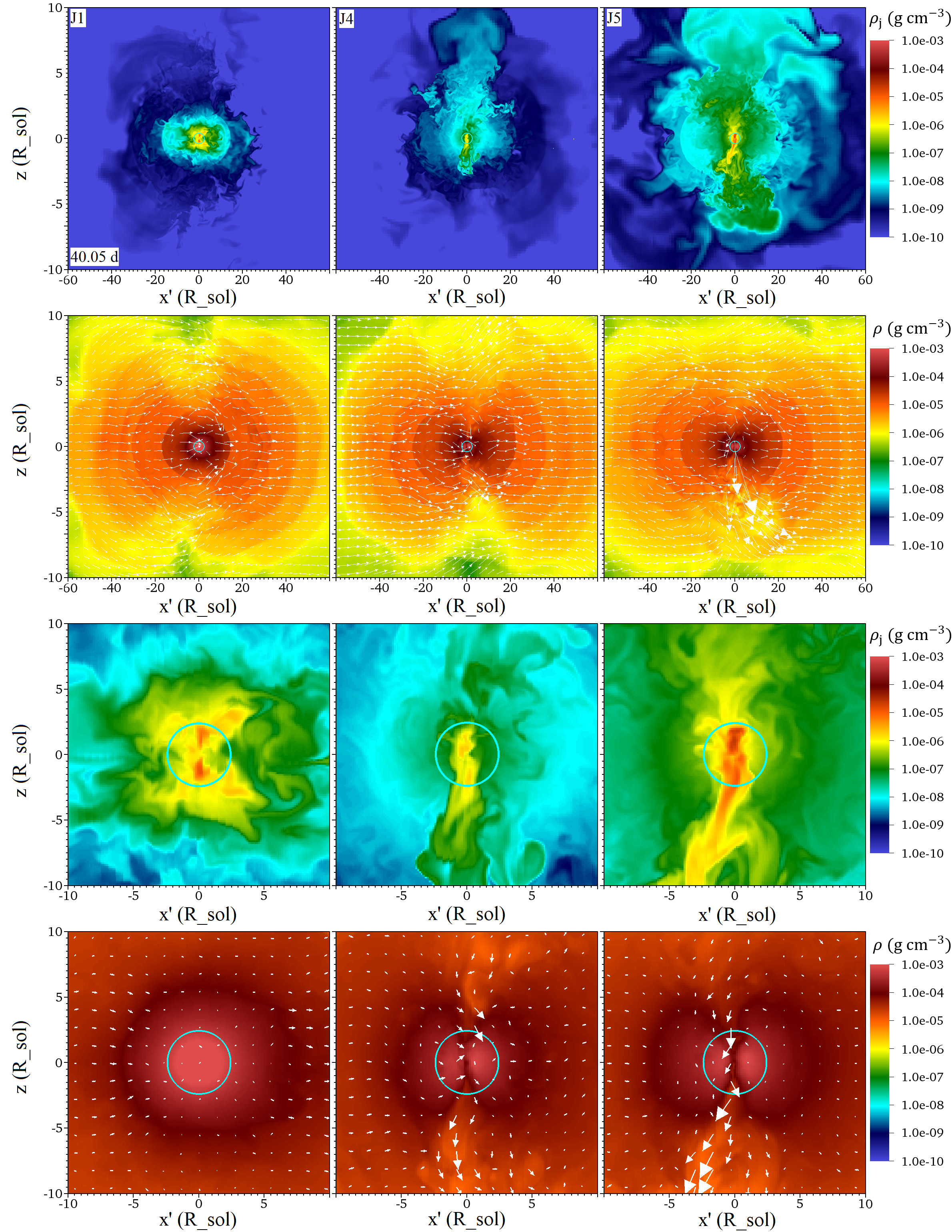}
    \caption{Slices at the final time $t=40.0\da$, taken in the same view as the first two rows in Fig.~\ref{fig:rhojet},
    show, from top to bottom rows, 
    (i) tracer density, (ii) gas density with velocity vectors in the rest frame of the secondary, 
    (iii) tracer density (zoomed in) and (iv) gas density (zoomed in with velocity vectors in the rest frame of the secondary).
    \textit{Left column:} Run J1 (fiducial, no subgrid accretion). 
    \textit{Middle column:} Run J4 (As J1 but with subgrid accretion turned on). 
    \textit{Right column:} Run J5 (As J4 but with jet mass-loss rate $\dot{M}\jet$ increased by a factor of $10$). 
    In the runs with subgrid accretion, jet gas has extended above and below the orbital plane (first and third rows).
    This is caused by entrainment of thermalized jet gas by envelope gas flowing through the funnel-shaped channel
    formed by the (highly super-Eddington) accretion, after the jet has choked. 
    Note that jet gas remains subdominant in mass (and energy) to co-spatial envelope gas in J4 ($\dot{M}\jet=\dot{M}\Edd$), 
    which implies that the jet has \textit{not} broken out. 
    In J5 ($\dot{M}\jet=10\dot{M}\Edd$) 
    the jet gas has \textit{comparable} density and energy to co-spatial envelope gas --
    since jet gas does not significantly \textit{displace} envelope gas, we do not consider this to be jet breakout.
    }
    \label{fig:accretion-jet}
\end{figure*}

Fig.~\ref{fig:rhojet} shows the density of \textit{only} that part of the gas which was injected by our jet subgrid model
(the jet tracer gas), for our fiducial run J1.
The top two rows show slices through the secondary perpendicular to the orbital plane
and to the 
%imaginary 
line joining the two particles,
with the direction of the primary core particle out of the page.
The secondary is at the centre of the frame with softening radius labeled by a circle, 
and the component of its motion parallel to the slice is toward the left.
The bottom two rows show slices through both particles and perpendicular to the orbital plane,
with the primary core situated to the left of the secondary and its softening sphere shown with a magenta circle.
The evolution of the orbital separation between the particles $a(t)$ is shown in the top panel of Fig.~\ref{fig:separation},
with Run~J1 represented by a solid green line.

By $t=1.1\da$, the jet has propagated out to $\sim\pm60\Rsun$ from the secondary. 
By $t=5.8\da$, as the secondary plunges into the envelope, the jet begins to lose coherence and connectivity.
Then, by $t=10.4\da$, the jet has acquired a ``<'' morphology (with $\phi$-component of its motion toward the left;
see the third panel in the top row).
To explain this, consider that the vertical component of the jet velocity has reduced
owing to the resistance provided by envelope material in and around the jet launch region.
Further, jet material is being entrained, to some extent,
by envelope gas which moves at a speed of order $v\orb$ with respect to the secondary.
As the two speeds become comparable, the jet acquires a significant angle with respect to the vertical,
and this angle increases with time. 

By $t=12.7\da$, around the time of the first periastron passage,
the angle of the ``<'' is smaller
and the jet is less extended and narrower due to its confinement by surrounding gas.
Evidence of Kelvin-Helmholtz instability -- caused by the relative motion between the jet and envelope material -- is visible.
The jet then oscillates between ``<'' and ``>'' morphologies as the secondary 
dives into dense material and re-emerges during the second periastron passage at $t\approx17.5\da$.
There is quite a bit of symmetry between the upper and lower portions of the jet,
but at $t=10.4\da$ significant asymmetry is present and the top portion is more deformed than the bottom.
Similar asymmetry when the jet begins to be choked is seen in other runs as well,
e.g.~the angles made with the vertical by the top and bottom portions of the jet can be quite different.
By the end of the simulation, the jet material is confined to a quasi-spherical region. 
Inside of this region, the structure of the jet is still somewhat bi-polar,
as seen in the right panel in the second row of Fig.~\ref{fig:rhojet}.
The jet has been completely choked by the surrounding envelope gas.
The energy it continues to eject is rapidly thermalized. 
After the first periastron passage, jet material added to the grid remains bound to the secondary
and has Mach number $\mathcal{M}<1$.
Other runs show an overall similar choking of the jet.

%\subsubsection{Influence of the jet on the envelope and vice-versa}
%\label{sec:gas}
To further study jet choking and how the jet affects envelope gas,
we plot in Fig.~\ref{fig:compare-J1-NJ1} snapshots of the total gas density (including jet, envelope and ambient gas)
overlaid with velocity vectors in the rest frame of the secondary, 
for runs with and without a jet. 
Both cuts of Fig.~\ref{fig:rhojet} 
(perpendicular to the line joining the particles and through the particles)
are shown, with Run~J1 in the first and third columns and Run~NJ1 in the second and fourth columns.
Times are the same as the first four snapshots of Fig.~\ref{fig:rhojet},
and the last snapshot approximately coincides with the first periastron passage.

At first, the jet is prominent in the snapshots
and  drills through the envelope by displacing envelope gas.
By the first periastron passage, however, the J1 and NJ1 snapshots are almost indistinguishable.
At $t=10.4\da$, shocks caused by the marginally choked jet are visible on either side of the orbital plane
in the slice through both particles.
The shocks are much more visible below the orbital plane in the figure,
and it is apparent from the velocity vectors that the lower half of the jet is choked a bit later than the top half.

In tandem, these various results show that the jet is choked around the time of first periastron passage,
in broad agreement with the theoretical estimates presented in Section~\ref{sec:choke}.

\subsubsection{Role of accretion and lack of jet breakout}
\label{sec:accretion}
To understand how accretion can affect the results, we compare Runs~J1 and J4,
which are identical except that J4 also allows accretion as in Model~B of \citet{Chamandy+18}.
Accretion rates after the first periastron passage are in the range $\sim(0.2$--$2)\Msunyr$ \citep{Chamandy+18}. 
This is equal to $\sim(100$--$1000)\dot{M}\Edd$ for an MS star and $\sim(10^4$--$10^5)\dot{M}\Edd$ for a WD.
The Eddington value always assumes spherically symmetric flow.
While it can be exceeded by a factor of $\sim10$ in axially symmetric flows, 
we consider the rates quoted above to be upper limits.
We thus study how maximally strong accretion affects CEE that includes a jet,
keeping in mind that smaller accretion rates might be expected to result in the same effects, only weaker.

In Fig.~\ref{fig:accretion-jet} we present various snapshots of density at the end of the simulation 
at $t=40.0\da$ for fiducial run J1 (left column), subgrid accretion run J4 (middle) 
and $10$-fold higher jet mass, subgrid accretion run J5 (right).
From top to bottom, we have the density of jet material alone (as in Fig.~\ref{fig:rhojet}), 
total gas density (as in Fig.~\ref{fig:compare-J1-NJ1}), 
density of jet material (zoomed in) and total gas density (zoomed in).
To facilitate direct comparison between the panels, the same colour table is used throughout.

Up until $t\approx30\da$, the morphologies of the jets in J1 and J4 are quite similar. 
At about this time, the blob of jet material in J4 starts to expand upward. 
Just before the simulation ends, the jet tracer gas suddenly reverses direction and  begins to extend downward. 
Jet gas remaining from the first episode is visible above the secondary, 
and the gas involved in the ongoing episode is visible below the secondary in Fig.~\ref{fig:accretion-jet}.
The jet gas in J4 is able to expand in the polar directions because of the channel 
that forms when the accretion subgrid model is active \citep{Chamandy+18}.
The polar regions become relatively depleted of envelope gas, 
providing a channel through which jet gas can flow.
Comparing the density of jet material with the total gas density 
we see that the density is dominated, as in J1, by \textit{envelope} material even where the jet material is located.
As the temperature and velocity are the same for co-spatial jet gas and envelope gas in our simulation,
the envelope gas dominates \textit{energetically} as well.
We thus conclude that this behaviour is caused by entrainment of thermalized jet material by envelope material 
flowing through the double-funnel-shaped partially evacuated region inside the torus.

While some of the material entering the subgrid accretion sphere of radius $4\delta_4\approx0.56\Rsun$ accretes, 
much of the gas flowing toward the torus centre instead passes through to the other side.
The same behaviour is seen in simulations with accretion but no jet,
and even in those simulations, the flow switches directions seemingly at random.\footnote{
This behaviour was observed in the runs with subgrid accretion of \citet{Chamandy+18},
but is being reported here for the first time.}
It is not clear how physical this particular flow pattern is,
and greater fidelity in this aspect would require more sophisticated modeling of the (unresolved) secondary
\citep[e.g.][]{Prust20}.
Because the jet is completely dominated by envelope gas,
we do \textit{not} consider this to be a true breakout of the jet.
On the other hand, if $\dot{M}\jet$ is increased by a factor of $10$, as in Run~J5,
the jet density is comparable to the density of envelope gas in the polar regions,
as seen in the right column of Fig.~\ref{fig:accretion-jet}.
Even in this extreme case, however,
the jet does \textit{not} truly break out because it does not significantly displace envelope material,
and jet gas barely 
%shows up in 
% \luke{
contributes to
% } 
the total gas density.
Other runs show less or no entrainment of jet gas by envelope gas along the polar directions.
%In any case, 
% \luke{
Therefore, true jet breakout never occurs in our runs.
% }
% \luke{
Moreover, the
% } 
high accretion rates 
% \luke{
which are found to promote entrainment of jet material
% }
represent an upper limit,
and it remains to be seen whether this behaviour would still occur 
if more realistic accretion rates for an MS or WD ($\lesssim10\dot{M}\Edd$) were obtained or assumed.

On the other hand, if the jet continues to inject energy at the same rate up until the envelope ejects, 
then at \textit{some point} before that the jet 
%must actually 
would naturally break out.
With this in mind, we can conclude from the lack of jet breakout in our simulations 
that jet breakout could only happen well after the $10$-orbit mark, 
and 
%even 
much later still 
for more realistic accretion rates 
% \luke{
at least
% } 
$1$--$2$ orders of magnitude smaller than those obtained.
% \luke{
The later the jet breaks out, 
the longer it stays coupled to the envelope, enabling efficient energy transfer that assists unbinding.
Therefore, longer simulations which can help to constrain the time of jet breakout would be valuable.
% }

\subsection{Orbital evolution}
\label{sec:orbit}
The evolution of the orbital separation $a(t)$ for all of the runs involving a $M_2(0)=0.978\Msun$ companion is shown in Fig.~\ref{fig:separation}.
We plot the runs without subgrid accretion in the top panel and those with it in the bottom panel,
with NJ1 repeated in both panels for ease of comparison.

In the top panel, we see that all runs including a jet but not subgrid accretion 
end up with slightly smaller orbital periods than the run without a jet.
Our fiducial run, J1, is more than half a period out of phase with NJ1 by $t=40\da$,
and the mean separation is correspondingly smaller.
However, the majority of this difference is apparently caused by the mass loss $\dot{M}_2=-\dot{M}\jet$,
since the orbit in Run~J3, which is identical to Run~J1 but with $\dot{M}_2=0$, is closer to that of NJ1 than to that of J1.
It is known that making the companion mass smaller results (for equal initial separation) in an initially longer period,
followed by a shorter period, 
and that the smaller the companion mass, the later the transition happens \citep{Passy+12a,Chamandy+19b},
and this likely explains the difference between J1 and J3.

While the dynamical friction drag on the secondary would be slightly enhanced by jet material that remains near to it,
it is not clear whether this explains why J3 ends up with a slightly smaller period than NJ1.
Doubling the jet speed, in going from J1 to J6, 
leads to a slight increase in the period, implying weaker drag. 
This may be because less of the jet material remains bound.
Run J7, which is like J1 except that the jet opening angle is doubled, 
also has a larger orbital period compared to J1,
perhaps because the wider jet is better at preventing envelope material from accumulating near the secondary.

\begin{figure}
    \centering
    \includegraphics[width=\columnwidth
    ]{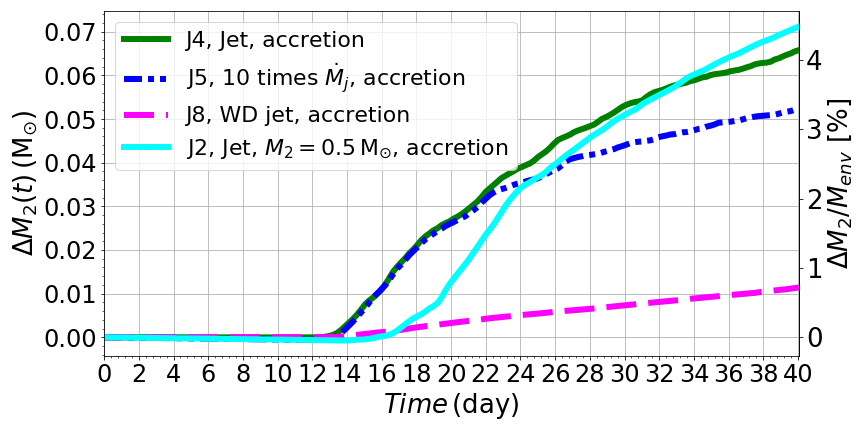}
    \caption{
    Change in mass of the secondary $\Delta M_2(t)= M_2(t) - M_2(0)$ 
    for the runs with subgrid accretion turned on.
    The accretion model is the same as that used in \citet{Chamandy+18},
    except that the accretion rate in Run~J8 has been capped at a constant value.
    In addition to mass gain due to accretion, 
    $\Delta M_2$ depends on a small constant rate of mass loss to supply the jet (see Tab.~\ref{tab:runs}).
    The right vertical axis shows $\Delta M_2$ divided by the initial envelope mass.}
    \label{fig:accretionrate}
\end{figure}

We now turn to the runs with accretion, shown in the bottom panel of Fig.~\ref{fig:separation}.
All runs with accretion lengthen the orbital period compared to those without accretion.
This is consistent with the known result (mentioned above) that more massive secondaries result in longer periods (eventually).
But another cause might be that accretion clears away material near the secondary, reducing drag.
Increasing $\dot{M}\jet$ by an order of magnitude from J4 to J5 causes the period to first be longer,
and eventually to be shorter, with the transition occurring at $t\approx35\da$.
This seems consistent with the companion accreting somewhat less mass after $t\approx22\da$ in J5 than in J4,
as seen in Fig.~\ref{fig:accretionrate}, 
though the reason that a larger $\dot{M}\jet$ leads to a dip in the accreted mass is not immediately clear.

\begin{figure}
    \centering
    \includegraphics[width=\columnwidth]{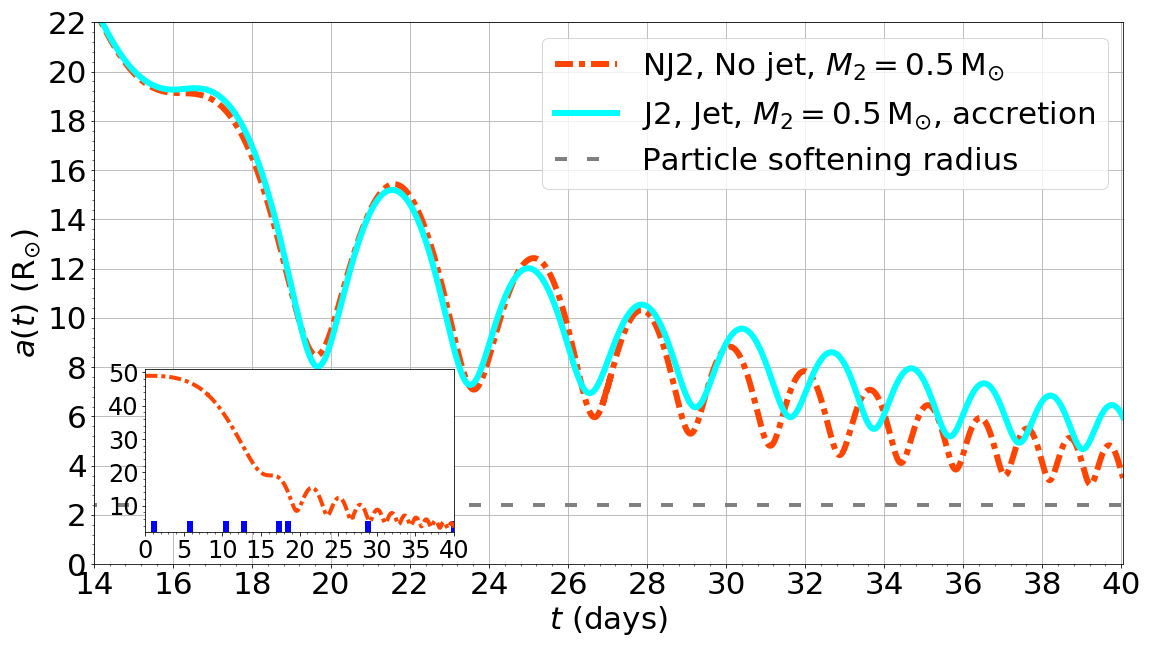} % ,clip=true,trim= 70 15 100 70
    \caption{Evolution of the inter-particle separation in the orbital plane for Runs~J2 and NJ2, 
    which have initial companion mass $M_2(0)=0.489\Msun$, i.e.~half of that used in the other runs. 
    }
    \label{fig:separation-group3}
\end{figure}

The WD run J8 has a separation curve with period only marginally longer than NJ1;
the shorter period compared to J4 or J5 could be due to the smaller accretion rate.
Finally, we turn to runs J2 and NJ2, which are similar to J5 and NJ1, respectively, 
but with companion mass only half as large.
These runs are plotted in Fig.~\ref{fig:separation-group3}.
J2 has a somewhat longer orbital period than NJ2 by the end of the simulation,
likely due to rapid accretion by the companion.

\subsection{Envelope unbinding}
\label{sec:unbinding}

\begin{figure}
    \centering
    \includegraphics[width=\columnwidth,clip=true,trim= 60 60 20 40]{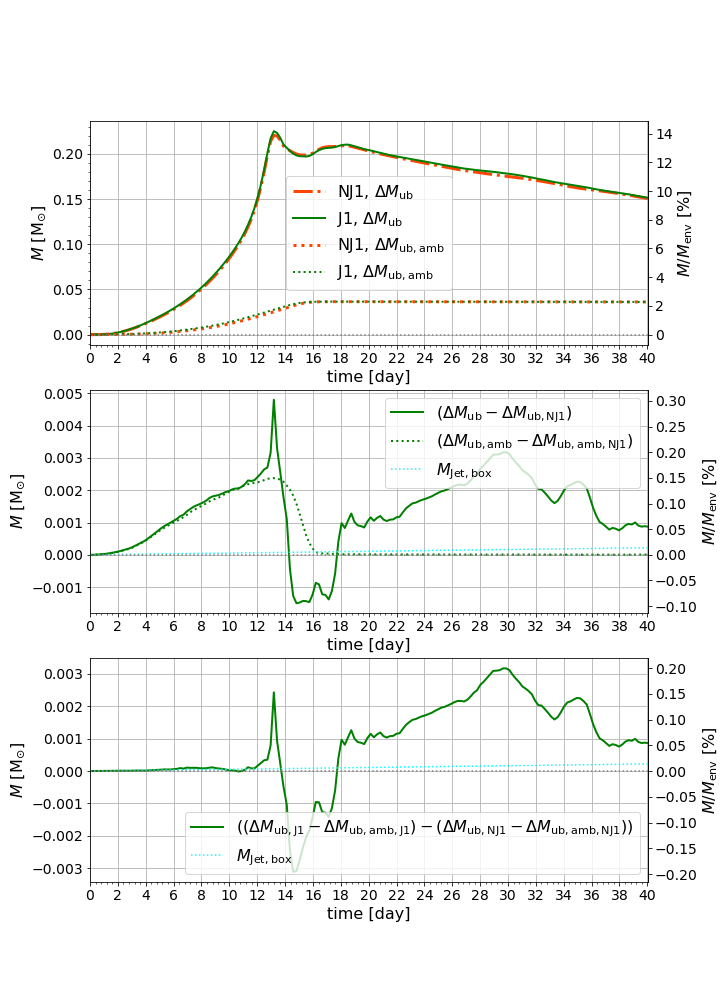}
    \caption{
    \textit{Top:} 
    Mass of all unbound gas (envelope, jet, ambient) relative to the initial value, 
    for Run J1 (fiducial jet model) and Run NJ1 (fiducial model without jet).
    Unbound mass of the ambient gas alone is also shown.
    Note that the unbound mass expressed as a percentage of the initial envelope mass is shown on the right vertical axis.
    \textit{Middle:} 
    Difference in unbound mass between simulations J1 and NJ1, shown for all unbound gas (solid) and ambient gas (dotted).
    The mass of jet material is relatively small and is shown as a dotted line.
    \textit{Bottom:} 
    Difference in the unbound mass between Runs J1 and NJ1,
    including envelope and jet material but excluding ambient material.
    At $t=40.0\da$, J1 has unbound an extra $\sim10^{-3}\Msun$ of envelope material, 
    i.e.~an extra $\sim1\%$ compared to the total unbound envelope mass of $\sim0.11\Msun$
    ($=0.15\Msun-0.04\Msun$ from the top panel).
    }
    \label{fig:unbound-J1}
\end{figure}

\begin{figure*}
    \centering
    \includegraphics[width=0.95\textwidth,clip=true,trim=20 35 0 20]{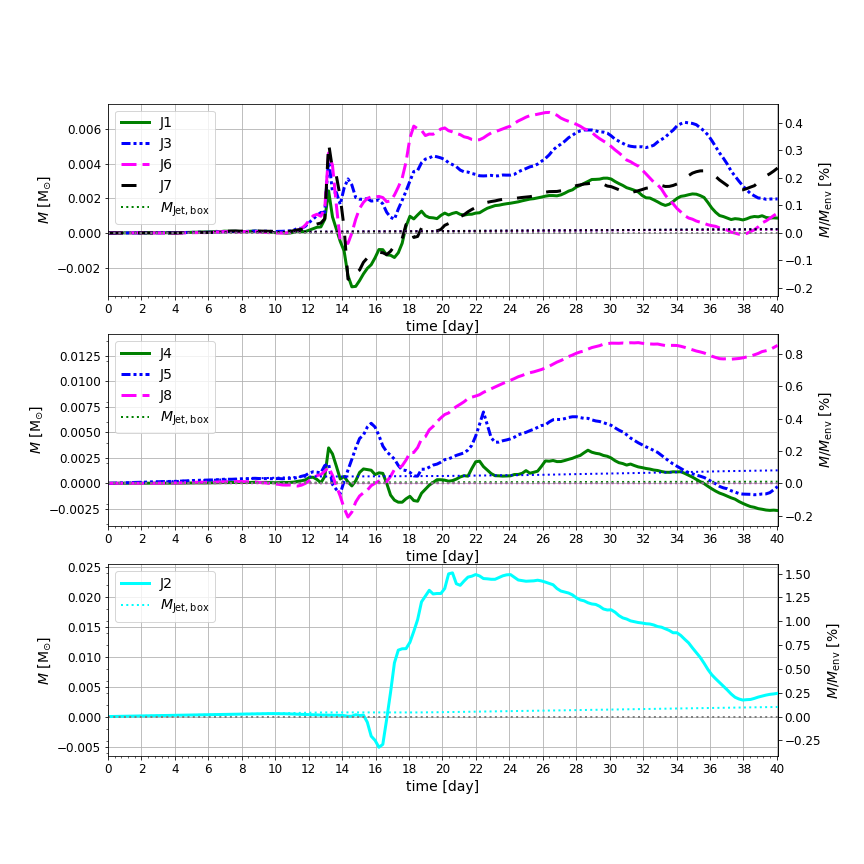} % 
    \caption{
    Same as bottom panel of Fig.~\ref{fig:unbound-J1}, now for all of the runs.
    \textit{Top:} Runs without subgrid accretion.
    \textit{Middle:} Runs with subgrid accretion involving a $\sim1\Msun$ secondary.
    \textit{Bottom:} Run J1, involving a $\sim1/Msun$ secondary, as compared to its no-jet counterpart.
    In each plot, dotted lines represent the accumulated jet mass $\dot{M}\jet t$ (see Tab.~\ref{tab:runs}).
    }
    \label{fig:unbound-envelope-all}
\end{figure*}

\subsubsection{Definition}
\label{sec:def}
We designate gas as ``unbound'' if $\En_{\mathrm{kin,gas}} + \En_{\mathrm{int,gas}} + \En_{\mathrm{pot, gas-gas}} + 2\En_{\mathrm{pot, gas-1}} + 2\En_{\mathrm{pot, gas-2}}>0$, where the terms are respectively
the kinetic energy density of bulk motions, 
the internal energy density, the potential energy density due to gas self-gravity,
twice the potential energy density due to the gravitational interaction between gas and the RGB core particle,
and twice the potential energy density due to the gravitational interaction between gas and the secondary 
(see \citealt{Chamandy+20} for details).
We choose to include the factors of two in the last two terms 
because for the gas and cores to ultimately unbind from one another,
the \textit{total} potential energy must be balanced, 
including the half nominally contained in the core particles \citep[e.g.~\S2.1 of][]{Binney+Tremaine08}.
No diagnostic predicts with certainty whether a given gas parcel will eventually be ejected,
so any such choice is somewhat arbitrary, and the choice used here is fairly conservative 
(for a discussion see \citealt{Ivanova+13a}).

\subsubsection{Extra unbinding caused by jets}
\label{sec:unbinding_jet}
The difference in the mass of unbound gas with respect to the simulation initial condition $\Delta M\unb(t)
\equiv M\unb(t)-M\unb(0)$
is plotted in Fig.~\ref{fig:unbound-J1} for Runs~J1 and NJ1.
Here $M\unb(0)$ is the $\sim98\%$ of ambient gas mass that is already unbound at $t=0$.
Mass flux through the boundaries of the grid is fully accounted for in the plots.
The top panel shows $\Delta M\unb$ for all the gas, as well as for ambient gas alone.
Note that all of the ambient gas that is bound at $t=0$ (found closest to the binary)
is rendered unbound by $t\approx16\da$ in each simulation.

About $12\%$ of the envelope mass is unbound by the first periastron passage 
(obtained by subtracting the unbound ambient gas mass from the unbound gas mass).
Thereafter, the unbound mass plateaus and slowly decreases.
This decrease is inconsequential for the analysis below
because we are interested in differences in the unbound mass between the runs.%
\footnote{The decrease may be caused by energy
transfer from marginally unbound gas causing it to become marginally bound,
subtle effects stemming from our chosen definition for ``unbound''
or energy transfer to the ambient medium.
We have explored other definitions 
(factors of $2$ omitted or thermal energy excluded) and still find a decrease with time.
This is consistent with the results of \citet{Prust+Chang19} 
(with initial conditions almost identical to our own),
who find a decrease until about $150\da$ 
using a definition of ``unbound'' that excludes our factors of $2$ and thermal energy.
However, such a decrease is not seen in \citet{Ohlmann16}, 
who employs similar initial conditions.}

The middle panel shows the difference in $\Delta M\unb$ between J1 and NJ1,
i.e.~the excess unbound mass caused by the jet.
Finally, the bottom panel subtracts the ambient mass and therefore shows the excess
unbound \textit{envelope} mass in J1 compared to NJ1.%
\footnote{To be precise, this quantity also includes the unbound jet mass. 
However, the total jet mass (cyan dotted line in Fig.~\ref{fig:unbound-J1}) is relatively small, 
implying that its contribution to the unbound mass is fairly negligible.}
Note that the right vertical axis shows the unbound mass as a percentage 
of the initial envelope mass $M\env(0)\approx1.6\Msun$.

We can measure the relative contribution of the jet to envelope unbinding
by dividing the \textit{excess} mass of unbound envelope gas in the jet run by the mass of all unbound envelope gas. 
At the end of our simulations at $t=40.0\da$, 
this amounts to $\sim10^{-3}\Msun/(0.15\Msun-0.04\Msun)\sim1\%$, 
where we have used the top panel and subtracted the unbound ambient mass 
from the unbound gas mass to obtain the unbound envelope mass.
Therefore, the jet causes an additional $\sim1\%$ of envelope mass to be unbound by the end of Run~J1,
as compared to Run~NJ1.
Prior to the end of the simulation, the excess unbound mass in the jet simulation can be a few times higher,
but between the first and second periastron passages it is actually negative 
(i.e.~less mass is unbound in J1 than NJ1 at that time).
Taking the simulation as a whole, 
including a jet leads to a \textit{small increase} in unbound envelope mass.

In Fig.~\ref{fig:unbound-envelope-all}, 
we plot the relative difference in the unbound mass compared to the corresponding no-jet run 
(as in the bottom panel of Fig.~\ref{fig:unbound-J1}) for all of the jet runs.
The top (middle) panel includes runs without (with) subgrid accretion,
while the bottom panel shows Run~J2 (smaller secondary mass) relative to Run~NJ2.

In all jet runs, 
the peak unbound mass, which occurs near the first periastron passage,
is higher as compared to the corresponding no-jet run.
Jets are generally seen to enhance the unbound envelope mass by a few percent. 
The excess unbound mass caused by the jet peaks at some point after $t=26\da$ for runs involving a $\sim1\Msun$ secondary,
but sometimes decreasing fairly sharply after peaking.
There are two cases where the excess unbound mass caused by the jet is significantly higher.
In Run~J2, the peak unbound envelope mass,
which occurs at $t\approx20\da$, near the first periastron passage,
is about $8\%$ of the initial envelope mass.
This is about $20\%$ higher than in Run~NJ2.
However, by the end of the simulation the extra unbinding in J2 compared to NJ2 is marginal.
This is probably because the final orbit of J2 is much wider than that of NJ2 owing to accretion (Section~\ref{sec:orbit}),
resulting in less liberation of orbital energy.
In Run~J8, our WD run, almost an additional $1\%$ of the envelope mass, 
or $\sim10\%$ of the \textit{unbound} envelope mass,
is unbound by the end of the simulation, as compared to NJ1.

%Therefore, 
In summary, we generally find that
simulations with jets (or jets+accretion) lead to between about 
$\sim1\%$ and $\sim10\%$ 
of extra unbinding compared to identical simulations without jets (or jets+accretion).

\subsubsection{Parameter space exploration}
Here we compare the various jet runs to better understand the effects of the model parameters.
Run~J3 is the only run for which the jet mass is not subtracted from that of the secondary,
and is otherwise identical to Run~J1.
We see from the top panel of Fig.~\ref{fig:unbound-envelope-all}
that J3 unbinds about twice as much extra envelope mass (compared to NJ1) as does J1.
This must be caused by the larger mass of the secondary.
Note that a dip to negative values of the excess unbound envelope mass 
immediately following the first periastron passage at $t\approx13\da$ is seen in all runs except J3.
This tells us that this brief period of \textit{reduction} in the unbound envelope mass as compared with 
the corresponding no-jet run is at least partly caused by the reduction of the secondary mass as it feeds the jet
(in Runs J2, J4, J5 and J8, accretion eventually produces a net increase in $M_2$).

Run~J6 is like J1 except with twice larger jet speed $v\jet$.
This results in a $4$-fold higher rate of kinetic energy release into the jet.
Consistent with this higher jet power, the excess unbound envelope mass is a few times larger than in J1, 
but, for reasons not yet understood, decreases at the end of the simulation to be comparable to J1.

Run~J7 is like J1 except with a twice larger jet opening angle.
The unbinding curve is similar to that of J1, 
but shows slightly more unbound mass at the first periastron passage.
This may be due to a larger surface of interaction between the jet and envelope material
as the secondary plunges in.

On the whole, turning on subgrid accretion does not have a large effect on the unbound mass.
This can be seen by comparing the surplus unbound mass in J1 (Fig.~\ref{fig:unbound-envelope-all} top panel, green solid)
with that in J4 (bottom panel green solid).
Therefore, differences in unbound mass between accretion+jet runs and no-accretion/no-jet runs 
can primarily be attributed to the jets.

The wider orbits of the runs with subgrid accretion imply a smaller amount of orbital energy transfer to the envelope. 
This results in lower unbound mass at late times, 
and likely explains the dip to negative values of excess unbound mass at the end of Runs~J4 and J5.
The jet in J5 is about $10$ times as powerful compared to that in J4 due to its $10$ times larger $\dot{M}\jet$.
The peak value of excess unbound envelope mass in J5 is more than twice that in J4. 
The tendency for a greater jet power to lead to more unbound material is compensated
somewhat by the lower secondary mass $M_2(t)$ in J5 
(due to feeding the jet at a higher rate and to a smaller accretion rate compared to J4).

Run~J2 is the run that produces the highest relative increase in the unbound mass at around the first periastron passage 
(peak of the unbound envelope mass curve) at $t\approx20\da$.
The jet power is generally larger as compared to the rate of orbital energy release
in J2 compared to the other jet runs.
In addition, envelope material near the jet has less binding energy due to the smaller secondary mass.
Finally, the smaller secondary mass implies less potential energy added by the jet.
However, by the end of the simulation, the wider orbit (due to accretion) negates any gain in unbound mass due to the jet.
With a smaller, more realistic accretion rate, extra unbinding at a level of $\sim20\%$ could perhaps 
be sustained for much longer.

The WD run J8 is the only case that exhibits a $\sim10\%$ increase in the unbound mass compared to NJ1 
by the end of the simulation.
Here $\dot{M}\jet$ is $100$ times smaller than in J1, so a comparison with J3 is more apt.
The peak excess unbound mass is about twice as high in J8 as compared to J3.
The kinetic energy of the jet is the same in both runs due to the $10$ times larger 
%$v\jet$ 
jet speed in J8.
Once the jet is choked, this kinetic energy transfers to heat in the envelope.
However, while the kinetic energy input is similar, 
J8 injects $100$ times less negative potential energy.
This might explain the greater unbinding in J8 as compared to J3.
Since J8 is the only run for which the net energy injected by the jet, including potential energy, is positive,
we consider this jet model to be the most realistic 
(though the potential energy injected by the jet can be thought of as sourced from the core particle).

\section{Overall impact of jets on envelope unbinding}
\label{sec:disc_unb}
In Section~\ref{sec:energy} we estimated the ratio of the kinetic energy supplied by the jet
to the orbital energy released during inspiral and found this ratio to be equal to about $(0.1$--$1)\%$.
Sure enough, including a jet in our simulations did not produce a drastic difference in the unbound envelope mass. 
However, it produced an increase in the unbound mass of $\sim(1$--$10)\%$.
Run~J8, which simulates a WD companion, 
is particularly noteworthy because it unbinds an extra $\sim10\%$ of envelope mass compared to Run~NJ1,
even though the estimated energy ratio is $\sim0.1\%$.
Thus, we conclude that jets may cause a fractional increase in unbound mass higher than their fractional energy input. 
This is likely because the jet energy is being distributed  efficiently to unbind already marginally 
bound gas, or prevent marginally unbound gas from becoming bound again.

%Importantly, the rate of energy supplied by the jet is only about $\dot{M}\Edd v\jet^2/40$
%owing to the angular jet profile used \citep{Federrath+14},
%but this choice of profile was somewhat arbitrary.
%If the angular profile were flat, then the rate of energy input would be about $10$ times higher.
%If, further, $\dot{M}\jet$ were increased from $\dot{M}\Edd$ in J8 to $10\dot{M}\Edd$ 
%(a reasonable upper limit) as in some of the other runs,
% \luke{
More extreme MS or WD jets with up to two orders of magnitude larger powers are perhaps possible.
% }
%then the jet power would increase by two orders of magnitude in total.
%might have somewhat flatter velocity profiles
%and higher mass loss rates, which would results in
We 
%argued 
estimated in Section~\ref{sec:tjet} that 
%jets $20$ to $200$ times more powerful than those used in our simulations
such powerful jets
might \textit{alone} unbind the envelope within $\sim10\yr$, 
comparable to the CE timescale estimated based on extrapolation from simulations that did not include jets.
Therefore, it seems plausible that maximally powerful MS or WD jets
($1$--$2$ orders of magnitude more powerful than those simulated)
could cause an $\mathcal{O}(1)$ increase in the rate of envelope mass unbinding,
and additional simulations are needed to investigate this possibility
% \luke{
(though even if such powerful jets could exist, 
it is doubtful whether they could be sustained for the duration of the CE phase).
% }

We find that the \textit{excess} unbound mass caused by including the jet spikes around the time of first periastron passage.
This is also around the time that the jet becomes choked and the \textit{overall} unbound envelope mass peaks 
(in simulations with or without a jet).
However, the excess unbound mass generally increases to its maximum values some time \textit{after} this 
(Fig.~\ref{fig:unbound-envelope-all}).
From the beginning of the simulation up until about the first periastron passage,
the jet interacts only weakly with the envelope for two reasons.
First, the tidal stream that emanates from the envelope and wraps around the secondary 
is concentrated at small polar angles relative to the secondary, where the jet is not present or weak
(Fig.~\ref{fig:compare-J1-NJ1}; 
this effect is enhanced by our choice of a strongly peaked angular jet velocity profile).
Second, the jet drills through the low density material of the outer envelope very easily,
i.e., without transferring much of its energy.
We can estimate the fraction of jet energy that does work on the surrounding gas
to be of order $\sim t\vel/t\orb$, where $t\vel$ is given by equation~\eqref{tv} and $t\orb$ by equation~\eqref{torb}
\citep[c.f.][]{Soker16}.
This gives an efficiency of about $(1$--$2)\%$, assuming $a=(47$--$47.5)\Rsun$, 
corresponding to $t\approx(4.5$--$5)\da$ and $v\orb\approx100\kms$.
However, some of this energy would be transferred to already unbound gas,
so the actual efficiency would be $\lesssim1\%$.
%$<1\%$.
On the other hand, once the jet chokes (and the energy it releases is thermalized locally), 
virtually all of its energy is transferred to bound envelope gas surrounding the secondary.
%This allows the choked jet to ultimately play a more important role in envelope unbinding compared than the unchoked jet.
% \luke{
Therefore, the ability of the jet to unbind efficiently envelope material
relies on the jet remaining choked.
Once it breaks out (which would happen eventually if it did not turn off first),
its effect on further envelope unbinding would likely be marginal.
% }

\section{Comparison with previous studies}
\label{sec:literature}
\subsection{``Grazing envelope evolution''}
``Grazing envelope evolution'' (GEE; \citealt{Soker15}) is a proposed scenario
where the jet unbinds envelope material and thus prevents it from accumulating 
around the secondary and causing dynamical friction drag.%
\footnote{An apparent contradiction entailed by this scenario is that the secondary would need to be accreting envelope gas
in order to power the jet.
% \luke{
This potential problem could perhaps be circumvented if the accretion disc acquires enough mass
before jet activation to sustain accretion after it, but more work is needed.
% }
}
%Nevertheless, a
Assuming that the jet can remain powered at a steady rate (as in our simulations),
could the CE phase be prevented by ``grazing''?
%\lukec{Moved some text to a footnote and added a sentence.}

GEE is not seen in our simulations nor in those of \citet{Shiber+19}, 
which are, to our knowledge, 
the only global simulations to include all the relevant gravitational forces as well as jets.
To test the viability of the GEE scenario, our simulations would ideally begin with a larger initial separation.
However, \citet{Shiber+19} did use a large initial separation for some of their runs,
and still did not report seeing GEE.
If GEE were common, one might expect a difference in the orbital evolution
between jet and no-jet runs at very early times.
But the separation curves of our jet and no-jet runs are extremely similar 
up to the first periastron passage (Figs.~\ref{fig:separation} and \ref{fig:separation-group3}).

One reason we do not see evidence of GEE could be that the
efficiency with which the jet unbinds material before it chokes is very small,
as discussed in Section~\ref{sec:disc_unb}.
However, a smaller mass companion, larger initial separation, 
or jet that is more powerful, wider or inclined relative to the vertical
may enhance the likelihood of a GEE-like scenario unfolding.

\subsection{Previous CE jet simulations}
\label{sec:other_work}
The global simulations by \citet{Shiber+19} had a similar setup to ours,
and were run for a similar number of orbits.
As in the present work, they found that jets choke as they becomes surrounded by dense envelope gas.
Just prior to choking they develop a ``<'' morphology 
(in the plane perpendicular to the orbital plane and 
%imaginary 
line joining the particles) similarly to what we found.
Unlike us, however, they found that the choked jet eventually breaks out.
On the other hand, \citet{Lopez-camara+21} found that even the strongest jets they simulate
choke before the first periastron passage and do not break out for the remainder of the simulation.
These results agree qualitatively with our own (though the parameter values of the systems simulated are very different).
However, it should be noted that all of the \citet{Lopez-camara+21} simulations 
end before the first pronounced periastron passage, 
which occurs just prior to $70\da$ (see simulation\#5 of \citealt{Shiber+19}).

The jet model of \citet{Shiber+19} removes material from their conical jet initialization regions at rates similar
to the highly super-Eddington accretion rates ($\gtrsim100\dot{M}\Edd$) that result from our accretion subgrid model.
In our simulations, jet breakout does not occur
but jet gas does expand in the polar directions in runs that allow accretion,
owing to the partially evacuated funnel-shaped regions that form on either side of the orbital plane
and entrainment of jet gas by envelope gas flowing through this region.
Therefore, it seems likely that the jet breakout seen in simulations of \citet{Shiber+19} 
is enabled by their mass removal prescription,  
which (as they explain) is somewhat akin to an accretion subgrid model.
As in our simulations which allow accretion, 
they find that the orbital period eventually becomes longer when jets are included.

When jet gas re-emerges from the dense quasi-spherical concentration of gas around the secondary in our simulations with accretion,
the jet material remains energetically subdominant compared to the envelope gas in its immediate surroundings,
and thus the jet does not truly break out (Section~\ref{sec:accretion}).
This is not the case in \citet{Shiber+19}, 
where the low-density jet is able to displace envelope material (i.e.~break out) in at least some cases.

While accretion in our simulations happens within $4\delta_4\approx0.56\Rsun$ from the secondary,
\citet{Shiber+19} continuously expunge \textit{all} envelope material within the jet cones, 
which extend out to $7\Rsun$ or $14\Rsun$, depending on the simulation.
Furthermore, they pressurize their jet by giving it thermal energy density comparable to that of the envelope gas removed.
As thermal energy of the envelope gas around the secondary likely almost balances gravitational potential energy
\citep{Chamandy+18}, this choice likely further facilitates jet breakout.
Buoyancy of the dilute jet gas 
%would 
could also help the jet to break out.
For these reasons, the jet breakout observed in \citet{Shiber+19} may not be realistic. 

Comparing similar simulations with and without a jet,
\citet{Shiber+19} concluded that the presence of a jet increased the outward flux of material 
through a sphere of radius $1\au$ or $2\au$ 
centred on the origin of the simulation domain and extending out to the boundary.
The fraction of the material which had positive energy density flowing out of this sphere was also 
determined to be higher when a jet was included
(but note the different simulation durations of the jet and no-jet runs).
Based on this they claim that jets unbind roughly three times as much envelope mass 
as identical simulations without jets.
Firstly, this result may depend sensitively on the ability of the jet to break out,
which, as argued above, may rely on questionable subgrid prescriptions.
Secondly, they did not state whether the fraction of the \textit{total} gas mass 
that acquired a positive energy density --
a more direct measure of unbinding  -- was different for jet and corresponding no-jet runs. 
As such, they did not actually 
determine whether the jets caused more material to become unbound
or only changed the spatial distributions of bound and unbound material.

\subsection{Neutron star and black hole secondaries}
\label{sec:NS}
NSs and BHs can undergo so-called hypercritical accretion,
with an accretion rate several orders of magnitude above the Eddington rate.
A NS jet with $\dot{M}\jet=0.032\Msunyr$ (about $10^6\dot{M}\Edd$ if $R_2=11.1\km$) and jet velocity $v\jet=0.1c$ 
(at a distance of $r\jet/2\approx1.1\Rsun$ from the NS)
would have a power of $\sim9\times10^{43}\ergs$, 
which is several orders of magnitude more powerful than the jets in our simulations.
This would also greatly exceed the mean rate of orbital energy release of $\sim3\times10^{40}\ergs$ in the simulations presented.
Using equation~\eqref{rho_choke} with the fiducial values $r\jet=2.25\Rsun$, $\theta\half=15^\circ$ 
and $M_2(0)=0.978\Msun$,
we find that the jet would choke when the envelope density near the secondary is $\sim3.7\times10^{-3}\gcmcmcm$,
which occurs at a radius $\sim0.8\Rsun$ in the initial envelope profile.
This is comparable to the final separation predicted using the $\alpha\CE$ energy formalism \citep{Chamandy+19a},
so the jet may or may not choke during the CE phase.
If not, then our results suggest that this may reduce its capacity to unbind the envelope.

To take another example, the same jet but now with $\dot{M}\jet=10^4\dot{M}\Edd$
instead of $10^6\dot{M}\Edd$ is predicted to choke at a density of $\sim3.7\times10^{-5}\gcmcmcm$,
which occurs at $26\Rsun$ from the centre in the initial RGB profile.
In this case, the jet would likely choke at around the first periastron passage.
Subsequently, if it continued to supply energy at the same rate,
its power would exceed the rate of orbital energy release by a factor of a few,
and the jet would thus likely dominate the envelope unbinding.
However, it might not remain choked and quickly break out,
reducing the efficiency of energy transfer to the envelope.
In any case, these rough examples 
%show 
suggest 
%that 
there may be a region of the CE parameter space involving NS and BH secondaries where jets 
dominate envelope unbinding \citep[e.g.][]{Hillel+21}.
% \luke{
Further studies involving global CE simulations with NS or BH jets are needed.
% }

\section{Conclusions}\label{sec:conclusions}
We simulated a common envelope phase involving a $2\Msun$, $48\Rsun$ RGB primary
and a $1\Msun$ or $0.5\Msun$ secondary that continuously launches a jet with approximately constant power.
The jet was included using a subgrid model that adds high-velocity gas 
to two spherical sectors on either side of the orbital plane 
% \luke{
(see Section~\ref{sec:jet_model}, Appendix~\ref{sec:jet_details} and
% }
\citealt{Federrath+14}). % at a constant rate.
We explored jet opening angles of $\theta\half=(15$--$30)^\circ$ 
but the jet velocity was strongly peaked inside $\theta\half/6$.
Our jets injected kinetic energy equal to $\sim(0.025-0.25)\dot{M}\Edd v\jet^2$,
with $\dot{M}\Edd$ the Eddington accretion rate for an MS star or WD 
and $v\jet$ the peak ($\theta=0$) jet speed, chosen to be of order (slightly larger than) the escape speed.
The results of our jet runs were compared with identical runs without a jet,
and we performed several runs to explore the parameter space for MS and WD secondaries.
Some of our runs included subgrid accretion onto the secondary 
at rates that we consider to be upper limits \citep{Krumholz+04,Chamandy+18},
but the contribution of the jet to envelope unbinding was found 
%to be insensitive 
not to be very sensitive 
to whether or not subgrid accretion was turned on.
The duration of our simulations was about $10$ orbits.
Our main results can be summarized as follows:
\begin{itemize}
  \item Jets in the simulations get choked 
  %at or shortly after 
  at around the time of the first periastron passage,
  and remain choked for the duration of each simulation,
  in broad agreement with rough analytic estimates and recent ``wind-tunnel'' type simulations 
  which used input from global simulations \citep{Lopez-camara+21}; 
  \item After choking, jets can enhance envelope unbinding by depositing energy locally,
  whereas before choking, jets tend to interact relatively weakly with envelope gas, 
  hardly affecting envelope unbinding;
  %\luke{Hence, transfer of jet energy to unbind envelope mass would likely becomes much less efficient if the jet breaks out.}
  \item Jets can 
  %contribute importantly to envelope unbinding \luke{in the simulations}, 
  enhance the unbound envelope mass in the simulations by up to about $10\%$,
  as for our simulation involving a WD companion (compare the curve representing J8 in Fig.~\ref{fig:unbound-envelope-all} 
  with the top panel of Fig.~\ref{fig:unbound-J1});
  \item The extra fractional unbound mass caused by jets during a simulation 
  can far exceed the fractional increase in the energy added, 
  likely because the jets can energize and unbind  marginally bound gas 
  or prevent marginally unbound gas from rebinding;
  \item Based on analytic estimates and rough extrapolation of simulation results
  %\textit{may}
  there remains a possibility that over the course of the full CE phase, 
  MS or WD jets might, in the most optimistic cases,
  contribute to envelope unbinding at a level comparable to that resulting from orbital inspiral alone
  (Sections~\ref{sec:tjet} and \ref{sec:disc_unb}).
  But this would likely require extremely powerful jets that remain choked, yet strongly active, 
  over much of the CE phase;
  \item Qualitative disagreement between our results and those of \citet{Shiber+19},
  who find that jets can break out after choking and greatly enhance envelope unbinding,
  %can 
  may be attributable
  %ed 
  to two 
  %likely 
  sources:
%   \lukec{I reversed the order}
  (i)~their jet subgrid model artificially promotes
  jet breakout by continually removing all envelope gas in the jet launch region and replacing it 
  with jet material with pressure equal to that of the material it replaces,
  and 
  (ii)~the measure they used for envelope unbinding does not take into account \textit{all} of the envelope gas; 
  %\item Our result that jets choke 
  %and do not break out (before the end of the simulation)
  %is consistent with the findings of \citet{Lopez-camara+21}, 
  %who ran local versions of one of the \citet{Shiber+19} simulations
  %with more realistic subgrid accretion+jet models and higher resolution;
  \item We see no evidence whatsoever for a GEE phase \citep{Soker15},
  but we cannot 
  %rule out 
  exclude the possibility of this type of scenario occurring 
  for cases with higher jet speeds, lower mass companions, larger initial separations,
  initial rotation of the primary, larger jet opening angles,
  and jets which have axes that are inclined relative to the orbital axis 
  \citep[e.g.][]{Schreier+19};
  \item We estimate that for CEE involving an NS or BH companion launching a jet,
  the jet could play a more important, 
  possibly dominant role in envelope unbinding because these jets could be much more powerful.
\end{itemize}

%Exploring a more extreme, 
%but still plausible model with a jet launched by an MS star or WD that is $10$--$100$ times as powerful as in our simulations
%%by using a flatter jet velocity angular profile 
%is one avenue for future work.
Future studies should 
% \luke{
involve longer simulations at higher resolution,
% },
tie the accretion rate to the jet mass-loss rate
(perhaps keeping the fraction of accreted mass that goes into the jet fixed)
% \luke{
and explore maximally powerful MS and WD companion jets.
% }.
Similar simulations with NS or BH companions would also be interesting.
However, such cases are challenging owing to the small timesteps necessitated by simulating high jet speeds
(to say nothing of modeling relativistic effects).

\section*{Acknowledgements}
This work used the computational and visualization resources in the Center for Integrated Research Computing (CIRC) at the University of Rochester. 
%and the computational resources of the Texas Advanced Computing Center (TACC) at The University of Texas at Austin
The authors acknowledge the Texas Advanced Computing Center (TACC) at The University of Texas at Austin for providing HPC resources that have contributed to the research results reported within this paper.
%, provided 
These were provided through allocation TG-AST120060 from the Extreme Science and Engineering Discovery Environment (XSEDE) \citep{xsede}, which is supported by National Science Foundation grant number ACI-1548562,
and through Frontera Pathways allocation AST20034, 
Financial support for this project was provided by the Department of Energy grants 
DE-SC0020432 and DE-SC0020434,
%\sout{DE-SC0001063}
the National Science Foundation grants 
AST-1813298 and PHY-2020249 (CMAP),
%\sout{AST-1515648, 
%and the Space Telescope Science Institute grant 
%HST-AR-14563.001-A. }
and the National Aeronautics and Space Administration grant 80NSSC20K0622.
%\lukec{Should we mention any of the commented out grants, which are no longer active?}
YZ acknowledges financial support from University of Rochester Frank J. Horton Graduate Research Fellowship.
%\lukec{Updated grant info from Bill Burroughs: GR507442 DOE DE-SC0020432,   GR507467 DOE DE-SC0020434,  GR507537 NASA 80NSSC20K0622, GR507129 NSF AST-1813298}

%%%%%%%%%%%%%%%%%%%%%%%%%%%%%%%%%%%%%%%%%%%%%%%%%%
\section*{Data Availability}
 
The data underlying this paper will be shared on reasonable request to the corresponding author.

%%%%%%%%%%%%%%%%%%%% REFERENCES %%%%%%%%%%%%%%%%%

\bibliographystyle{mnras}
\bibliography{refs}

\appendix

%\section{Plots that may not be needed}
%\begin{figure*}
%    \centering
%    \includegraphics[width=\textwidth]{Figures/1D-plots/accretion.png}
%    \caption{Another version of Fig. \ref{fig:accretionrate}. Top panel is the averaged accretion rate between frames. Bottom panel, the solid lines are the net accreted mass onto the companion $\dot{M}_2$, and the dotted lines are the total accretion by the Krumholz model $\dot{M}_a$. \amy{Need to adjust color code and line-styles.}}
%\end{figure*}

\section{Details of the jet subgrid model}
\label{sec:jet_details}

When injecting outflow mass, momentum, and angular momentum, it is desired to accurately conserve these quantities which would otherwise not be conserved due to the numerical discretization employed in subgrid models.  Here we solve for corrected values for mass and momentum injection that are close to the desired values while constrained by the conservation laws.  $\vec{R}$ and $\vec{V}$ are the particles position and velocity that remain unchanged following the injection of some amount of mass $M$, radial momentum $P_r$, and angular momentum $\vec{J}$.  We will also use primed quantities for those in the particles frame $\vec{r}' \equiv \vec{r} - \vec{R}$ and $\vec{v}' \equiv \vec{v} - \vec{V}$

\begin{itemize}
\item Mass conservation
\begin{equation}
  \qquad \displaystyle \sum m_i = M; \label{m}
\end{equation}
\item Conserving center of mass
\begin{equation}
  \qquad \displaystyle\sum m_i \vec{r}_i = M \vec{R} \rightarrow  \sum m_i \vec{r}'_i = 0; \label{CM}
\end{equation}
\item Momentum conservation
\begin{equation}
  \qquad \displaystyle\sum m_i \vec{v}_i = M \vec{V} \rightarrow \sum m_i \vec{v}_i' = 0; \label{p}
\end{equation}
\item Angular momentum conservation (with an additional source of angular momentum from the particles spin $\vec{J}$)
\begin{align}
\qquad  \displaystyle \sum m_i \vec{r_i} \times \vec{v_i} = M \vec{R} \times \vec{V} + \vec{J} \rightarrow \sum m_i  \vec{r}_i' \times \vec{v}_i' = \vec{J}   \label{J}
 \end{align}  Note this follows from \ref{m},  \ref{CM}, \& \ref{p}.
\item Desired amount of radial momentum in particle's frame
\begin{equation}
\qquad \displaystyle \sum_{i} m_i \vec{v}_i' \cdot \frac{\vec{r}_i'}{| \vec{r}_i' |} = P_r \rightarrow \sum_{i} m_i \vec{\scriptr}_i \cdot \vec{v}_i  = P_r \label{vi}
\end{equation}
where $\vec{\scriptr}_i \equiv \frac{\vec{r}_i'}{| \vec{r}_i' |}$.
\end{itemize}

Now writing $\vec{r}_i' \equiv [x_i, y_i, z_i]$ and $\vec{v}_i' \equiv [u_i,v_i,w_i]$ 
the first two constraints for total mass \ref{m} \& the center of mass \ref{CM} only depend on the mass and can be expressed as 

 \[
 \left[\begin{array}{rrrrr}
   1 & 1 & 1 & ... & 1  \\ 
  x_1 & x_2 & x_3 & ... & x_n  \\
  y_1 & y_2 & y_3 & ... & y_n  \\
  z_1 & z_2 & z_3 & ... & z_n  \\
   \end{array} \right]
 \left[\begin{array}{r}
  m_1  \\
  m_2  \\
  m_3  \\
  ...  \\
  m_n  \\
   \end{array} \right]=
 \left[\begin{array}{r}
  M  \\
  0  \\
  0  \\
  0  \\
   \end{array} \right]
\]

This system is in general under-determined since the kernel will have more points than constraints.  However, it  can be solved using a least squares approach, where we find the solution for $m_i$ that is as close as possible to our kernel function (i.e.~injection profile) for mass at those points $\mathcal{M}(\vec{r}_i')$ 

If we consider the above matrix equation as $Ax=b$ and represent the target kernel solution vector $d = \mathcal{M}(\vec{r}_i')$, then the solution $x$ closest to the target $d$ that satisfies the constraints is given by 

\begin{equation}
  x=d-A^T(AA^T)^{-1}(A d - b)
\end{equation}

Note this involves inverting the 4x4 matrix $(A A^T)$. 
Once we have solved for the masses, we can then solve for the velocities using the other 3 constraints 
(conservation of momentum \ref{p}, angular momentum \ref{J}, 
and desired scalar momentum injection \ref{vi} treating $m_i$ as knowns and $u_i$, $v_i$, and $w_i$ as unknowns.  
This gives us the matrix equation 
 
\tiny
 \[
 \left[\begin{array}{rrrrrrrrr}
  m_1 & ... & m_n & \zero & ... & \zero & \zero & ... & \zero \\
  \zero & ... & \zero & m_1 & ... & m_n & \zero & ... & \zero \\
  \zero & ... & \zero & \zero & ... & \zero & m_1 & ... & m_n  \\
   \zero & ... & \zero  & -m_1z_1 & ... & -m_nz_n &  m_1y_1 & ... & m_ny_n\\
  m_1z_1 & ... & m_nz_n &  \zero & ... & \zero & -m_1x_1 & ... & -m_nx_n \\
-m_1y_1 & ... & -m_ny_n &  m_1x_1 & ... & m_nx_n &\zero & ... & \zero  \\  
m_1 \scriptr_{1,1}& ... & m_n \scriptr_{n,1} &m_1 \scriptr_{1,2}& ... & m_n \scriptr_{n,2} &m_1 \scriptr_{1,3}& ... & m_n \scriptr_{n,3}  \\
   \end{array} \right]
   \]
   \[
 \left[\begin{array}{r}
  u_{1}  \\
  ...  \\
  u_{n} \\
  v_{1}  \\
  ...  \\
  v_{n} \\
  w_{1}  \\
  ...  \\
  w_{n} \\
    \end{array} \right]=
 \left[\begin{array}{r}
  0 \\
  0 \\
  0 \\
  J_x  \\
  J_y  \\
  J_z  \\
  P_r\\
   \end{array} \right]
\]

\normalsize
And again we can construct a target vector $d$ of velocities using our velocity kernel $\vec{\mathcal{V}}(\vec{r}_i')$ and solve for the velocities $x$ that satisfy the constraint while being as close as possible to the kernel.  

Note that it is important that our various kernels for mass, radial momentum, and angular momentum be somewhat close to satisfying the constraints to begin with.  The closer the solution is to satisfying the constraints, the 
smaller the resulting deviation from the target profile.  This is particularly important for the mass.  For instance, using an unnormalized mass kernel can lead to negative masses when solving the constraints.

%%%%%%%%%%%%%%%%%%%%%%%%%%%%%%%%%%%%%%%%%%%%%%%%%%

% Don't change these lines
\bsp	% typesetting comment
\label{lastpage}
\end{document}